\documentclass[11pt, a4paper]{article}
\usepackage[utf8]{inputenc}
\usepackage[T1]{fontenc}
\usepackage{authblk}
\usepackage[a4paper,top=2.5cm,bottom=2.5cm,left=2.5cm,right=2.5cm,marginparwidth=1.75cm]{geometry}

\usepackage[shortlabels]{enumitem}
\usepackage[ruled,vlined]{algorithm2e}

\usepackage[noend]{algpseudocode}
\usepackage{amsmath, amssymb}
\usepackage[numbers]{natbib}

\usepackage{setspace, array}
\usepackage{mathtools}
\usepackage{hyperref}

\usepackage{multirow}

\date{}

\author[1]{Osama Ahmed}
\author[1,2]{Felix Tennie}
\author[1,3-4,*]{Luca Magri}

\title{Robust quantum reservoir computers for forecasting chaotic dynamics: generalized synchronization and stability}

\affil[1]{Department of Aeronautics, Imperial College London, London SW7 2AZ, UK.}
\affil[2]{Department of Engineering, City St George's, University of London, London EC1V 0HB, UK.}
\affil[3]{The Alan Turing Institute, London NW1 2DB, UK.}
\affil[4]{DIMEAS, Politecnico di Torino, Torino 24 10129, Italy.}
\affil[*]{\textit{Corresponding author:} \texttt{l.magri@imperial.ac.uk}}

\date{}

\begin{document}

\maketitle

\begin{abstract}
We show that recurrent quantum reservoir computers (QRCs) and their recurrence-free architectures (RF-QRCs) are robust tools for learning and forecasting chaotic dynamics from time-series data. First, we formulate and interpret quantum reservoir computers as coupled dynamical systems, where the reservoir acts as a response system driven by training data; in other words, quantum reservoir computers are generalized-synchronization (GS) systems.
Second, we show that quantum reservoir computers can learn chaotic dynamics and their invariant properties, such as Lyapunov spectra, attractor dimensions, and geometric properties such as the covariant Lyapunov vectors. This analysis is enabled by deriving the Jacobian of the quantum reservoir update.
Third, by leveraging tools from generalized synchronization, we provide a method for designing robust quantum reservoir computers. We propose the  criterion $GS=ESP$: GS implies the echo state property (ESP), and vice versa. We analytically show that RF-QRCs, by design, fulfill $GS=ESP$.
Finally, we analyze the effect of simulated noise. We find that dissipation from noise enhances the robustness of quantum reservoir computers. Numerical verifications on systems of different dimensions support our conclusions. This work opens opportunities for designing robust quantum machines for chaotic time series forecasting on near-term quantum hardware.
\end{abstract}


\section{Introduction}

Chaotic dynamics \citep{lorenz_deterministic_1963,palmer_extended-range_1993} arise from different fields, such as astrophysics \citep{buchler2012chaos}, meteorology \citep{selvam2017nonlinear}, chemistry \citep{field1993chaos}, turbulence \citep{manneville2010instabilities}, thermoacoustics \citep{huhn2020learning}, 
among many others. The dynamics of chaotic systems can be modelled either with equation-based models or with data-driven models. Here, we focus on data-driven models. 
Predicting the temporal evolution of a chaotic dynamical system is challenging because infinitesimal errors and perturbations grow exponentially in time.
This makes the time-accurate prediction of chaotic systems challenging after the predictability time \citep{BOFFETTA2002367}, which is a characteristic scale of the physical system. On the other hand, chaotic solutions of ergodic systems can also be characterized by quantities, which are unaffected by infinitesimal perturbations; e.g., the statistics of the solution, and invariant properties such as the Lyapunov spectrum \citep{shimada1979numerical,benettin1980lyapunov}, the Kaplan-Yorke dimension \citep{frederickson1983liapunov}, 
covariant Lyapunov vectors \citep{ginelli2013covariant,margazoglou_stability_2023}, and  many more \citep{BOFFETTA2002367}. As argued in \citep{VKIlecturemagri,margazoglou_stability_2023,ozalp_reconstruction_2023,ozalp_stability_2024}, a data-driven model for chaotic time series is "good" when it 
(i) predicts  the evolution of the dynamics for a relatively long horizon in an autonomous way (i.e., after training), 
(ii)  predicts the long-term statistics of the solution, and 
(iii) infers the invariant properties of the solution. Key to the computations of most invariant properties is the Jacobian of the data-driven model. In the large variety of data-driven models that is available, we focus on reservoir computers (RCs) because their ans\"atz is principled and justified by  synchronization theory of chaotic systems \citep{pecora1990synchronization,pecora2015synchronization}. Synchronization in two chaotic systems can be achieved by designing a coupling between them. In the case of one-way coupling, a driving system (i.e., the physical system) provides the forcing to the response system, which is a higher dimensional system  \citep{pecora1991driving,badola1992driving}. The conditions for which the response system is synchronized with the driving system are provided by generalized synchronization (GS) theory \citep{rulkov1995generalized,kocarev1996generalized,yang1998generalized,abarbanel1996generalized}. Recent works in reservoir computers have shown that, when some conditions are met, classical RCs are  generalized synchronization systems with the training time series acting as the driving state and a reservoir state acting as the response state \citep{lu2018attractor,lymburn2019reservoir,platt2021robust}. Reservoir computers were employed to predict scalar invariant properties, e.g., the Lyapunov spectrum \citep{pathak2017using,vlachas2018data,vlachas_backpropagation_2020}, as well as geometric invariant properties, e.g.,  the covariant Lyapunov vectors \citep{margazoglou_stability_2023}, of the Lorenz system and the Kuramoto-Sivashinsky (KS) equation. Importantly, generalized synchronization is connected to the  echo state property (ESP), which  needs to be met for a reservoir computer's performance to be independent of the initial reservoir state, which, in turn, is necessary for a reservoir computer to forecast the time series  \citep{jaeger__2001,JAEGER2007335}. 
When the ESP is fulfilled, reservoir computers can also accurately infer other invariant quantities, such as how small changes to the physical parameters affect an objective functional for design optimization  \citep{ozan2024data}, and the chaotic properties in the latent space of autoencoders \citep{ozalp_stability_2024,paine_physics-informed_2023}. In this paper, we import the theory of GS into quantum reservoir computers.

Quantum mechanics offers ansatze for reservoir computers  \citep{fujii_harnessing_2017}.  Quantum reservoir computers  have been employed for time-series forecasting on classical \citep{Fujii2021,mujal_time_2023,pfeffer_hybrid_2022,ahmed2024prediction} and quantum data \citep{sornsaeng_quantum_2023} on near-term noisy quantum devices \citep{fry_optimizing_2023,domingo_taking_2023}. As compared to classical reservoir approaches \citep{lukosevicius_practical_2012,jaeger__2001,JAEGER2007335,racca_data-driven_2022}, a quantum benefit is achieved in terms of expressivity, i.e., the quantum substrate with entanglements  offers a rich feature generator to effectively learn nonlinear dynamics from data  \citep{kornjavca2024large,dudas_quantum_2023}. Because quantum reservoirs do not require backpropagation to be trained, they do not suffer from vanishing gradients, which are also known as \textit{barren plateaus} \citep{cerezo_variational_2021,mcclean2018barren,larocca2024review}. However, there are other challenges to be tackled such as the effect of sampling noise \citep{ahmed2025optimal,mujal_time_2023} and exponential concentration \citep{xiong2023fundamental}. 
The effect of finite sampling noise can  be reduced with  gate-based recurrence-free quantum reservoir computers (RF-QRCs)  \citep{ahmed2024prediction,ahmed2025optimal}, which are scalable and efficient machines, which make parametric studies computationally less demanding than they would be with recurrent quantum reservoir computer (QRCs). Recent works have proposed to redefine the ESP for quantum reservoir systems \citep{kobayashi2024extending,kobayashi2024coherence,martinez2023quantum}. Here, we propose an alternative  approach for defining  the quantum ESP by considering the quantum reservoir as a drive-response system, which requires GS.  Finally, the design of robust quantum reservoir computers also needs to consider noise. Noise can be projective (i.e., caused by finite sampling),  coherent and incoherent  \citep{9781107002173}. The effect of noise  was investigated in \citep{sannia_dissipation_2022,sannia2024engineered,domingo_taking_2023,fry2023optimizing}, which show that noise enhance the  performance. In this work, we analyse the performance in noisy environments from a GS perspective. 

The overarching goal of this paper is to make a connection between generalized synchronization theory and quantum reservoir computers to enable the design of robust machines in both noise-free and noisy environments.  QRCs and RF-QRCs are "robust" when 
(i) they fulfill the echo state property;
(ii) they predict the chaotic time series beyond the predictability time over a range of initial conditions;
(iii) they correctly infer the invariant properties of the chaotic solution; and 
(iv) the range of hyperparameters for which their performance is "good" is large.

The paper is structured as follows. In Section~\ref{sec:learning}, we introduce the concept of generalized synchronization (GS). In Section~\ref{sec:QRC}, we make the connection  between GS and the learnability of chaotic systems with quantum reservoir computing. In Section~\ref{sec:stab_qrc}, we analytically derive the Jacobians of QRCs and RF-QRCs, and deploy them to infer the invariant properties of  chaotic systems from data. In Section~\ref{sec:prac_qrc}, we provide  practical guidelines for designing robust quantum reservoir computers  with GS both in noise-free and noisy scenarios. In Section~\ref{sec:conclusion}, we conclude the paper. Appendices provide further details. 

\section{Stability and generalized synchronization (GS) theory}\label{sec:learning}

In this section, we review  stability analysis of chaotic systems and generalized synchronization (GS) in drive-response systems. In Section~\ref{sec:QRC}, we show that quantum reservoir computers  are one-way coupled synchronized dynamical systems. This interpretation will enable us to exploit GS tools to analyse and design robust quantum reservoir computers  in Section~\ref{sec:prac_qrc}.  

\subsection{Chaotic systems and invariant properties from stability theory}\label{sec:stability}
We consider a nonlinear autonomous dynamical system
\begin{align}\label{eq:0}  
\frac{d}{dt}\tilde{\pmb{x}}(t) = f(\tilde{\pmb{x}}(t)), \,\,\,\, \tilde{\pmb{x}}(0) = \tilde{\pmb{x}}_0, 
\end{align}
where $\tilde{\pmb{x}}(t)\in  \mathbb{R}^D$ \footnote{The notation  $\tilde{\pmb{x}}(t)$ is simplified to $\tilde{\pmb{x}}$   unless it is required for clarity.}  is the state vector; $f:\, \mathbb{R}^{D} \rightarrow \mathbb{R}^{D}$ is a continuously differentiable nonlinear vector function; and $D$ is the number of degrees of freedom.  
In linear stability analysis, we analyse the evolution of an infinitesimal perturbation $ \varepsilon \pmb{w}$ applied to the state $\tilde{\pmb{x}}$ as
\begin{align}\label{eq:1}
\begin{split}    
\tilde{\pmb{x}} + \varepsilon \pmb{w}, \quad \quad \quad \varepsilon \rightarrow 0.   
\end{split}
\end{align}
Substituting Eq.~\ref{eq:1} into Eq.~\ref{eq:0} and retaining the first-order terms yield 
\begin{align}\label{eq:2}
\begin{split}    
\frac{d}{dt}\pmb{w}(t) = \tilde{\pmb{\textnormal{J}}}(\tilde{\pmb{x}}(t))\pmb{w}(t),  \,\,\,\, \pmb{w}(0) = \pmb{w}_0, 
\end{split}
\end{align}
where $\tilde{\pmb{\textnormal{J}}}$ is the Jacobian, $\tilde{J}_{ij}=\partial f_{i}/\partial \tilde{x}_{j}$ with $i,j=1,2,\ldots D$. (Eq. \eqref{eq:2} is known as the perturbation, or Jacobian, or tangent, or variational equation.) The perturbation, $\pmb{w}$, evolves in the tangent space spanned by the Jacobian, which is evaluated at the time-varying state, $\tilde{\pmb{x}}(t)$. The goal of stability analysis is to characterize the exponential growth and directions of infinitesimal perturbations. 
To do so, we numerically time march $D$ pseudorandom vectors, $\pmb{w}_{i}$, cast as columns of a matrix $\pmb{W}$, which is periodically QR-decompose as $\pmb{W} = \pmb{Q}\pmb{R}$, 
where $\pmb{Q}$ is an orthonormal matrix, and  $\pmb{R}$ is an upper-triangular matrix \citep{golub1996cf}. To compute stability properties, we follow the procedure outlined in \citep{huhn_stability_2020,margazoglou_stability_2023,ozalp2023reconstruction} (For completeness, the algorithm is shown in Appendix~\ref{app:algo}.) Oseledets multiplicative ergodic theorem \citep{oseledec1968multiplicative}, under mild assumptions, shows the existence of $D$ Lyapunov exponents $\lambda_{1} \geq \dots \geq \lambda_{D}$, which can be computed as
\begin{align}\label{eq:AA1}
\begin{split} 
\lambda_{i} = \lim_{T\to\infty} \frac{1}{T}  \int_{t_{0}}^{T} \ln[\pmb{R}(t)_{i,i}] dt.
\end{split}
\end{align}

The leading Lyapunov exponent $\lambda_{1}$ is the largest growth rate of the chaotic dynamics \citep{BOFFETTA2002367}. Almost every infinitesimal perturbation  grows exponentially as $||\pmb{w}(t)|| \sim e{^{\lambda_{1}t }||\pmb{w}(0)}||$ for $ t \rightarrow \infty$. If $\lambda_{1} < 0 $, the perturbations decay, i.e., the attractor is a fixed point. If  $\lambda_{1} = 0 $ the attractor is periodic (or quasi-periodic if there exist at least two neutral Lyapunov exponents), and if $\lambda_{1} > 0 $, the attractor is chaotic. The focus of this paper is on chaotic systems. The inverse of the leading Lyapunov exponent $\lambda_{1}$, also known as the Lyapunov time (LT), is used in this paper to scale the physical time units. 
The dominant portion of the Lyapunov spectrum provides the Kaplan-Yorke dimension, which is an upper bound of the attractor dimension \citep{frederickson1983liapunov}
\begin{align}\label{eq:12}
\begin{split} 
D_{KY} = l + \frac{ \sum_{i=1}^{l} \lambda_{i}}{|\lambda_{l+1}|},
\end{split}
\end{align}
in which $l$ is defined as  $\sum_{i=1}^{l} \lambda_{i} > 0$ and $\sum_{i=1}^{l+1} \lambda_{i} < 0$. \\

The Lyapunov spectrum and the Kaplan-Yorke dimension are scalar invariants (measures) of the chaotic attractor.
Geometric invariants, which describe the vector structure of the tangent space, are the covariant Lyapunov vectors (CLVs) \citep{ginelli2013covariant,kuptsov_theory_2012,ginelli_characterizing_nodate}. 
CLVs, by definition, are covariant with the dynamics and invariant under time reversal. CLVs provide a (generally) non-orthogonal, local splitting of the tangent space into unstable, neutral, and stable subspaces, corresponding to positive, zero, and negative Lyapunov exponents, respectively. The (absolute) angle between  pairs $\pmb{v}_{i}, \pmb{v}_{j}$ of CLVs is 
\begin{align}\label{eq:11}
\begin{split} 
\theta_{\pmb{v}_{i},\pmb{v}_{j}} := \frac{180}{\pi} \textnormal{cos}^{-1} (|\pmb{v}_{i} \cdot \pmb{v}_{j}|), \, \, \, [^{\circ}]\,\,\,\,\,\ i,j, = 1,2,\ldots, D
\end{split}
\end{align}
where ($\cdot$) denotes the dot product, provides information on the structure of the dynamical system (hyperbolic vs non-hyperbolic \citep{huhn_stability_2020}). In this paper, we use these angles as metrics to quantitatively assess the reservoir computers' performance in time forecasting the chaotic dynamics.

\begin{figure}[!h]
\centering\includegraphics[width=6in]{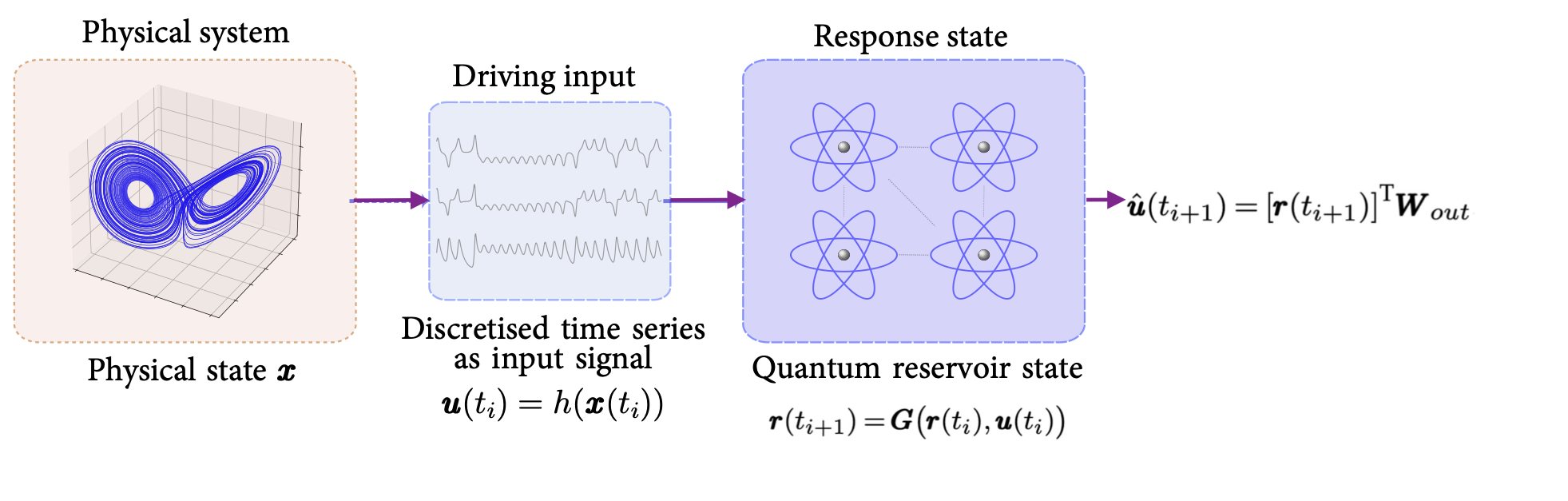}
\caption{Quantum reservoir computers as drive-response systems. The chaotic time series (data), $\pmb{x}$, is available at discrete intervals over $[t, t+1, ..]$ as $\pmb{u}(t_i) =  h(\pmb{x}(t_i))$, with $h$ being the observable map. The driving input forces the quantum reservoir system $\pmb{r}$ (response system). The reservoir state evolves as \textsf{$ \pmb{r}({t_{i+1}}) = \pmb{G}\big({\pmb{r}(t_i),\pmb{u}(t_i)}\big))$} (Section~\ref{sec:QRC}). The linear map \textnormal{$\mathbf{W}_{out}$} is the only trainable part of the quantum reservoir computer.  }
\label{fig:GS}
\end{figure}

\subsection{Generalized synchronization (GS) and conditional Lyapunov exponents}\label{sec:GS_CLE}

We provide the background for  generalized synchronization (GS)  \citep{pecora1990synchronization,pecora1991driving,rulkov_generalized_1995,kocarev1996generalized,hunt1997differentiable,yang1998generalized, tang1998observation}.   %
Upon explicit time discretization, the continuous-in-time dynamical system in Eq.~\ref{eq:0} becomes a discrete map 
\begin{align}
    \pmb{x}(t_{i+1}) = \pmb{F}(\pmb{x}(t_i)), \,\,\,\, \pmb{x}_0 = \pmb{x}(0). 
\end{align}
The training time series, $\pmb{u}$, may, in general, be 
$\pmb{u}(t_i) =  h(\pmb{x}(t_i))$. In this paper, without loss of generality, we set $h$ to the identity operator, so $\pmb{u}(t_i) =  \pmb{x}(t_i)$. (This means that we focus on fully observable dynamical systems.) 
We consider a one-way coupled chaotic system, in which the driving state, $\pmb{u}(t_i)$,  one-way forces a response system described by the vector $\pmb{r}(t_i) \, \in \, \mathbb{R}^N$ with $N\gg D$. Figure~\ref{fig:GS} shows a  schematic of a one-way a coupled drive-response system with relevance to  quantum reservoir computers ({Section~\ref{sec:QRC}). The dynamics of the one-way coupled system are governed by
%
\begin{align}\label{eq:4c} 
\pmb{u}({t_{i+1}}) &=  
\pmb{F}\big(\pmb{u}(t_{i})\big), \,\,\,\, &\pmb{u}_{0} = \pmb{u}(0). \\
\pmb{r}({t_{i+1}}) &=  
\pmb{G}\big({\pmb{r}(t_i),\pmb{u}(t_i)}\big), \,\,\,\, &\pmb{r}_0 = \pmb{r}(0).\label{eq:4d}
\end{align}
where
 $\pmb{G}$ is a user-defined map, whose properties are discussed in the remainder of this section.   
The response state, $\pmb{r}$, is in generalized synchronization (GS) with the driving state, $\pmb{u}$, {when $\pmb{G}$ generates a  
map $\phi: \pmb{x} \rightarrow \pmb{r}$ that guarantees asymptotic stability}  (e.g., \citep{abarbanel1996generalized,hunt1997differentiable})  %
\begin{align}\label{eq:5}
\lim_{i\to\infty} \left \| \pmb{r}(t_i) - \phi (\pmb{x}(t)) \right \| = 0, \,\,\,\, \mathrm{for\,\, every}\,\, \pmb{x}(t) \in \mathcal{M}, 
\end{align}
where $\mathcal{M}$ is the attractor of the driving system (also referred to as synchronization \citep{verzelli_learn_2021,platt_systematic_2022} or physical manifold). 
In other words, GS occurs when there exists a continuous {$\phi$ that fulfills \eqref{eq:5}. Equivalently, GS occurs when    
i) physically, the response system converges to the  attractor, $\mathcal{M}$;
ii) mathematically, $\pmb{G}$ is a contractive map, i.e., it has a  Lyapunov spectrum smaller than unity; and 
iii) from a functional analysis point of view, 
$\pmb{r}(t_i)=\phi (\pmb{x}(t_i))$ is a Cauchy sequence that tends to $\phi (\pmb{x}(t))$ on the attractor $\mathcal{M}$.  
There might exist infinite maps, $\phi$, that guarantee GS; however, only a continuously differentiable {$\phi$} guarantees that the attractor $\mathcal{M}$ can be correctly reconstructed, thus, its invariant properties can be correctly inferred by the response system (differentiable generalized synchronization,  \citep{hunt1997differentiable}).
The Conditional Lyapunov Exponents (CLEs) of $\pmb{G}$ (Eq.~\ref{eq:4d}) and the Lyapunov exponents of $\pmb{F}$ (Eq.~\ref{eq:AA1}) determine the differentiability of $\phi$ \citep{pecora_driving_1991,pyragas1997conditional,hunt1997differentiable}. The CLEs are computed from the Jacobian of the response system  conditioned on the driving signal, $\pmb{u}$, as shown in Section \ref{f3uorqnfu432u34qhf7892hq34fh934f3q}.  
As  proved by \citep{hunt1997differentiable}, $\phi$ is continuously differentiable if condition (iii) in Table~\ref{tab: CLE_GS} is met. We will use this condition to design robust quantum reservoir computers. \\  

The relationship between Lyapunov exponents of  discrete maps ($\lambda_{\textrm{discrete}}$) and the Lyapunov exponents of the  corresponding continuous system ($\lambda_{\textrm{continuous}}$)  is 
\begin{align} \label{f4jw0f582j08f24}
\lambda_{\textrm{continuous}} = \frac{\ln(\lambda_{\textrm{discrete}})}{dt}, 
\end{align}
where $dt$ is the time step. 
In this paper, we show the Lyapunov exponents of the continuous system. 

\begin{table}[!h]
\centering
\caption{Criteria for GS and differentiability. The most negative Lyapunov exponent of the driving system is denoted $\lambda^*$. The relationship between the exponents of the continuous and discrete systems is provided by Eq. \ref{f4jw0f582j08f24}. }
\label{tab: CLE_GS}
\begin{tabular}{llll}
\hline
 & Condition &Condition  &  Property \\
&(continuous system)  & (discrete map)&  \\
\hline
i. &$\max(\lambda_{CLE}) \geq 0$ & $\max(\lambda_{CLE}) \geq 1$  & GS does not occur. \\ & & &\\
ii. & ${\lambda^*}< \max(\lambda_{CLE})<0$  &${\lambda^*}< \max(\lambda_{CLE})<1 $  & GS occurs   \\
& & &$\phi$ is not differentiable.\\ & & &\\
iii. &$ \max(\lambda_{CLE})  < {\lambda^*} <0 $ & $\max(\lambda_{CLE})  < {\lambda^*} <1$ & GS occurs \\ 
& & & $\phi$ is continuously differentiable. 
\\ & & & Differentiable GS (DGS).
\\ \hline

\end{tabular}
\vspace*{-4pt}
\end{table}

\section{Quantum reservoir computing}\label{sec:QRC}
We make a connection between quantum reservoir computing and generalized synchronization (GS).
In a quantum reservoir computing approach, a quantum system is the response system (reservoir)\footnote{From a hardware perspective, this quantum system can be an analog \citep{kornjavca2024large} or spin qubit system \citep{martinez2023information}, a superconducting quantum circuit \citep{suzuki_natural_2022}, transverse-field Ising model \citep{fujii_harnessing_2017}, or more generally a gate-based quantum circuit \citep{pfeffer_hybrid_2022,ahmed2024prediction}.}, and the driving states are the observation on a physical system, encoded in data. 
 The quantum ans\"atz with entangled qubits offers rich expressivity \citep{fujii_harnessing_2017, Fujii2021} to infer the mapping $\phi$ in Eq.~\ref{eq:5}. 
In gate-based quantum systems, the quantum reservoir state is a ket vector $|\psi \rangle$, which is propagated by a unitary $\mathcal{U}(\pmb{\theta})$ at each time step \citep{ahmed2024prediction}
\begin{align}\label{eq:3c1}
   {|\psi(t_{i+1}) \rangle } &=  \mathcal{U}(\pmb{\theta}) {|\psi(t_{i}) \rangle }, \\ \label{eq:3cc2}
    &= V(\pmb{\alpha}) \, \Xi(\pmb{u}(t_{i})) \, P(\pmb{r}(t_{i})) | 0 \rangle^{\otimes n}
\end{align}
where, $ V(\pmb{\alpha})$ is a  unitary with random parameters $\boldsymbol{\alpha} \in \mathbb{R}^{n}$, $\Xi(\pmb{u}(t_{i}))$ is a unitary that depends on the input data, $\pmb{u}$, and $P(\pmb{r}(t_{i}))$ encodes the effect of previous states for the reservoir (response) map as in Eq.~\ref{eq:4d}. The ket $| 0 \rangle^{\otimes n}$ is the tensorial product of the initial states of $n$ qubits. 

Each unitary evolution is made of classical input-dependent single-qubit rotations $R_{y}$ and two-qubit entanglement CNOT gates to encode and process the classical data for the quantum reservoir. Common choices for the unitary map, known as feature maps, were investigated in \citep{pfeffer_reduced-order_2023,ahmed2024prediction,suzuki_natural_2022}. In this work, we use a fully connected feature map for unitary evolution (Figure~\ref{fig:noise_cirq}) as in \citep{ahmed2024prediction}. After each time step, a measurement in the computational basis $\{|k\rangle\}_{k=0}^{k=2^n}$ is performed to  
form a new reservoir state vector $\pmb{r}(t_{i+1})$ \citep{ahmed2024prediction} 
\begin{align} \label{eq:3c3} 
{r}^{(k)}(t_{i+1}) = (1-\epsilon) \, r^{(k)}(t_{i}) + \epsilon \, |\langle \psi(t_{i+1}) | k\rangle| ^2 , 
\end{align}
where {
${r}^{(k)}$ is the component of the reservoir state on the basis vector $k$, 
}
and  $0\leq \epsilon \leq 1$ is the leak rate, which is a hyperparameter. In vector notation,
\begin{align}
\label{eq:3j8}
\pmb{r}(t_{i+1}) &= (1-\epsilon) \, \pmb{r}(t_{i}) + \ldots \\ \nonumber \ldots & \epsilon \, \langle 0 |^{\otimes n} \, \,    P^{\dag}(\pmb{r}(t_{i})) \, \Xi^{\dag}(\pmb{u}(t_{i})) V^{\dag}(\pmb{\alpha}) \,|k\rangle\langle k| \,  V(\pmb{\alpha}) \, \Xi(\pmb{u}(t_{i})) \, P(\pmb{r}(t_{i})) |0 \rangle^{\otimes n} , \\ 
\label{eq:3j9}
 &= (1-\epsilon) \, \pmb{r}(t_{i}) + \epsilon \, \langle 0 |^{\otimes n} \, \,    P^{\dag}(\pmb{r}(t_{i})) \, \Xi^{\dag}(\pmb{u}(t_{i})) \,\pmb{\mathcal{W}} \, \Xi(\pmb{u}(t_{i})) \, P(\pmb{r}(t_{i})) |0 \rangle^{\otimes n} , 
\end{align}
where  $\mathcal{W}^{(k)} = V^\dagger|k\rangle\langle k|V$ form a $2^n$-dimensional vector. Finally, the reservoir state is mapped back onto the physical domain as
\begin{align}
    \hat{\pmb{u}}(t_{i+1})= \pmb{r}(t_{i+1}){^{{\textnormal T}}} \pmb{W}_{out},    \label{f3rjqiwrjfi32qje3}
\end{align}
where $\hat{\pmb{u}}$ is the reservoir computer's prediction, and  $\pmb{W}_{out} \in \mathbb{R}^{{N}_{r} \, \times N_{u}}$ is a rectangular matrix, whose components are the only trainable parameters.
Quantum reservoir computers  are trained by minimising a quadratic error 
%
\begin{align}\label{eq:3de4}
E = || {\pmb{u}}(t_{i+1}) - \pmb{r}(t_{i+1}){^{{\textnormal T}}} \pmb{W}_{out} ||^{2} + \beta \,|| {\pmb{W}_{out}}||^{2}, 
\end{align}
where $\beta$ is the non-negative Tikhonov regularization factor, which is a hyperparameter. 
The training of quantum reservoir computers is a quadratic optimization problem (Eq.~\ref{eq:3de4}), whose global minimum is found by solving a linear system (ridge regression) 
\begin{align}\label{eq:3c4}
\begin{split} 
(\pmb{R}\pmb{R}^{T}+\beta\pmb{I}) \, \pmb{W}_{out} = \pmb{R} \, \pmb{U}^{{\textnormal T} }_{d},
\end{split}
\end{align}
where $\pmb{R} \equiv [\pmb{r}(t_{1}), \pmb{r}(t_{2}), ... \pmb{r}(t_{N_{tr}})] \in \mathbb{R}^{{N}_{r} \, \times N_{tr}}$ is a matrix with concatenated reservoir states corresponding to each neuron for $N_{tr}$ time steps of training, and $\pmb{U}_{d} \equiv [\pmb{u}(t_{1}), \pmb{u}(t_{2}), ... \pmb{u}(t_{N_{tr}})] \\ \in \mathbb{R}^{{N}_{u} \, \times N_{tr}}$ is the matrix of concatenated input time series (data). 
After training (open-loop phase) and validation, the reservoir can be deployed as an autonomous dynamical system (closed-loop phase) to predict the evolution of the physical state in the future. A "good" quantum reservoir computer accurately forecasts the time series and infers the invariant properties of the attractor. 
\begin{figure}[!h]
\centering\includegraphics[width=6in]{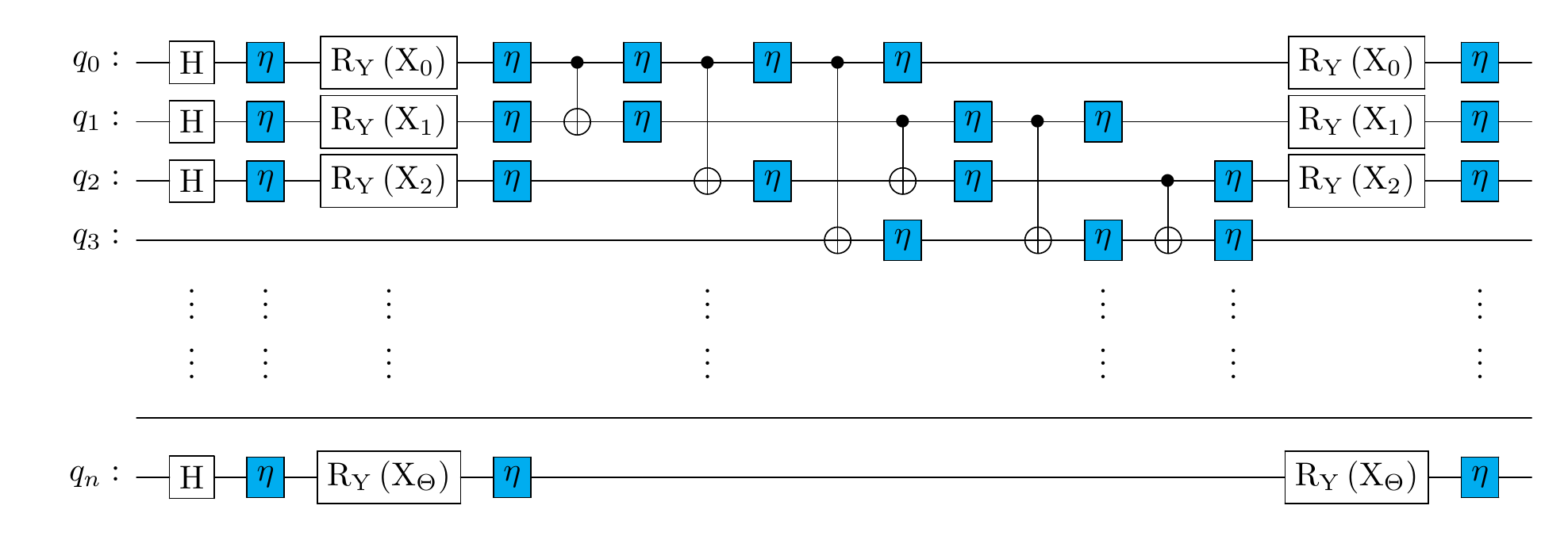}
\caption{Fully-connected quantum circuit for the RF-QRC \citep{ahmed2024prediction} architecture with noise channels $\eta$ after each gate operation (noise will be applied in Section \ref{sec:noise}). The number of encoded parameters $\Theta$ for RF-QRC is equal to the dimension of the input data ($\Theta = D$) for $\Xi({\pmb{u}}) $ unitary. When $D > n$, the feature map encoding is applied multiple times to encode all $\pmb{u}$ dimensions. For a random unitary $V({\pmb{u}})$, the number of encoded parameters are equal to the number of qubits $n$, ($\Theta = n$).  }
\label{fig:noise_cirq}
\end{figure}

\subsubsection{Two types of quantum reservoir computers}

 Quantum reservoir computers can be categorized as recurrent quantum reservoir computers, for brevity QRCs) or recurrence-free quantum reservoir computers (RF-QRCs). On the one hand,  QRCs  \citep{mujal_opportunities_2021,pfeffer_hybrid_2022} contain recurrences in the unitary $P(\pmb{r}(t_{i}))$ in Eq.~\ref{eq:3j9}. These recurrences keep memory of the past, which is key to time series forecasting. 
On the other hand, RF-QRCs  \citep{ahmed2024prediction,ahmed2025optimal}  set $P(\pmb{r}(t_{i})) = \pmb{I}$, where $\pmb{I}$ is the identity. The memory of the past is kept by leaky integration in Eq.~\ref{eq:3c3}. The leaky hyperparameter, $\epsilon$, determines the memory ($0<\epsilon\ll1$ preserves long memory, whereas the limit $\epsilon=1$ is that of  Extreme Learning Machines \citep{xiong2023fundamental}, which have no memory).
As shown in \citep{ahmed2024prediction,ahmed2025optimal}, RF-QRCs scale to higher-dimensional systems, are robust for chaotic time series forecasting, and require fewer tunable hyperparameters.
In this paper, both QRCs and RF-QRCs are analysed, but the focus in the main text is on RF-QRCs (Appendix~\ref{app:stab_recurrent} contains results on QRCs). 

\subsubsection{Echo state property}\label{sec:ESP}

For classical reservoir computers to perform well, it is necessary that the echo state property (ESP) is fulfilled   \citep{lukosevicius_practical_2012,JAEGER2007335,jaeger__2001,boedecker_information_2012,storm_constraints_2022}. 
The ESP is defined  as the property of a reservoir computer to "forget" the effect of initial conditions 
\citep{lu_attractor_2018,platt2021robust,platt_systematic_2022,hart2024attractor,farkas_computational_2016,racca_predicting_2023}. 
From a GS point of view, we argue that the echo state property is equivalent to the condition of GS in Eq.~\ref{eq:5}. GS and ESP are the same conditions under two different names, which originate from two different communities. In this paper, we propose the criterion $GS = ESP$, which is a shorthand to mean that GS implies ESP, and vice versa, in quantum computers.   

\begin{table}[!t]
\centering
\caption{Connection between GS and QRCs}
\label{tab: GS_eqv}
\begin{tabular}{lll}
\hline
& Generalized synchronization & Quantum reservoir computing \\ 
\hline
$\pmb{x}$ & Physical state (driving state)  & Physical state  \\
$h$ & Observable map $(\pmb{u} = h (\pmb{x}))$ & - \\
$\pmb{u}$ & Discrete driving state & Reservoirs input time series  \\
$\pmb{F}$ & Evolution map of driving system & - \\
\pmb{G} & Evolution map of response system & Evolution map of quantum reservoir computer \\
$\pmb{r}$ & Response state & Quantum reservoir state  \\
$\phi$ &  Defined implicitly by $\pmb{G}$ and $\pmb{F}$  &  $\phi^{-1} \equiv  \pmb{W}_{out}$ \\
Learnability & Asymptotic stability & Echo state property \\ \hline
\end{tabular}
\vspace*{-4pt}
\end{table}

\subsection{Casting QRC as a GS problem}

The reservoir computer ans\"atz \eqref{f3rjqiwrjfi32qje3} defines implicitly the local  inverse of the map $\phi$ in GS theory {(Section~\ref{sec:GS_CLE})}, i.e.,  $\phi^{-1}\equiv  \pmb{W}_{out}$. Drawing on classical reservoir computers,   \citep{lu_attractor_2018,platt2021robust,platt_systematic_2022,hart2024attractor, grigoryeva2021chaos}, we argue that the \textit{learnability} of a dynamical system can be achieved with a quantum reservoir computer if $\pmb{W}_{out}$ is the local inverse of a function that fulfills  at least {condition (ii) in Table~\ref{tab: CLE_GS}}, and, ideally, the stronger condition (iii). A good  $\pmb{W}_{out}$, which satisfies GS, results in an autonomous response system, which is deployed to forecast the time series and infer the invariant properties of the driving system. 
With tools from GS and dynamical systems, we propose two principles to  assess the performance and enable the design of quantum reservoir computers: 

\begin{itemize}
    \item During the training phase, the synchronization between two subsystems (drive-response) is assessed by computing the CLEs of the quantum reservoir update (Eq.~\ref{eq:3c3}), which is the one-way coupled drive-response system in GS theory (Eq.~\ref{eq:5qrca}). 
    \item In the autonomous evolution after training, the invariant properties of the physical manifold are computed using the Jacobian of the trained quantum reservoir system, which we analytically derive and show in section~\ref{sec:stab_qrc}. We use the tools in section~\ref{sec:stability} to compute these invariant properties, and the quantum reservoir state (Eq.~\ref{eq:3c3}) is the response system in GS theory (Eq.~\ref{eq:4d}). 
\end{itemize}

We will use these two principles   in Section~\ref{sec:prac_qrc} as a strategy to tune and design robust quantum reservoir computers. A summary on the connections between quantum reservoir computers and GS is shown in Table~\ref{tab: GS_eqv}.
A  hyperparameter tuning should fulfill at least condition (ii) in Table~\ref{tab: CLE_GS} to  guarantee the fulfillment of the ESP. 

\subsection{Contractive maps and GS in quantum reservoir computers}\label{sec:unitary}

The  map {$\pmb{G}$} in Eq.~\ref{eq:4d} ensures generalized synchronization when it is contractive   
\begin{align}\label{eq:3d1}
\left \| {\pmb{G}}(\pmb{r}_{1},\pmb{u}) - {\pmb{G}}(\pmb{r}_{2},\pmb{u}) \right \| \leq  \gamma \left \| \pmb{r}_{1} - \pmb{r}_{2} \right \|,         \
\end{align}
which means that, for the same driving state $\pmb{u}$,  two initial conditions, $\pmb{r}_{1} $ and $\pmb{r}_{2}$, converge to the same limit with  a rate  $0<\gamma<1$. This implies that the reservoir state $\pmb{r}$ asymptotically becomes independent of the initial state, thereby satisfying the ESP \citep{lu2018attractor}. 
In quantum reservoir computing, Eq.~\ref{eq:3d1} can be rewritten in terms of completely positive and trace-preserving (CPTP) quantum maps, $T$, which act on the quantum state represented by the density operator $\rho$ \citep{9781107002173}
\begin{align}\label{eq:3d2}
\begin{split} 
\left \| T({\rho}_{1},\pmb{u}) ) - T({\rho}_{2},\pmb{u}) ) \right \| \leq  \tilde{\gamma} \left \| {\rho}_{1} - {\rho}_{2} \right \|.
\end{split}
\end{align}

The map $T$ is unitary, therefore, $ \tilde{\gamma} = 1$. This is a fundamental property of quantum mechanics, which reflects the fact that unitary operations do not alter the relative geometrical configuration (e.g., rotation angles) between quantum states in Hilbert space. This means that unitary maps are non-expansive in trace norm. Therefore, the quantum reservoir state update shown in Eq.~\ref{eq:3cc2} (as is) does not satisfy $GS=ESP$ (Section~\ref{sec:ESP}). 
In this paper, we propose two methods to design a contractive quantum reservoir computer, which satisfy $GS=ESP$: 
(a) using leaky integration (Eq.~\ref{eq:3c3}) with $\epsilon < 1 $ to ensure a contractive update; and  
(b)  exploiting the dissipative nature of noise.
Either condition (a) or (b) makes 
the quantum channel in quantum reservoir computers strictly contractive ($\tilde{\gamma} < 1$ in Eq.~\ref{eq:3d2}). This is good news: In contrast to classical reservoir computers, which may have expansive maps that make the machine divergent \citep{huhn2022gradient,racca_statistical_2022}, we propose quantum reservoir computers that are contractive by design. (Further conditions on GS in quantum reservoir computers with injective generalized synchronization were recently explored in \citep{martinez2024input,martinez2023quantum}.)

\subsection{Jacobians}\label{f3uorqnfu432u34qhf7892hq34fh934f3q}
Key to computing the invariant properties of chaotic systems (section~\ref{sec:stability}), such as the Lyapunov exponents, CLVs, and the KY dimension,  is the Jacobian of the quantum reservoir computer update. 
The Jacobian of Eq.~\ref{eq:3c3} is 
\begin{align}\label{eq:9}
\hspace{-1cm}{\mathbf{J}}(\pmb{r}(t_{i}), \pmb{{u}}(t_{i})) = \frac{d\pmb{r}(t_{i+1})}{d\pmb{r}(t_{i})} =\frac{\partial \pmb{G}(\pmb{{u}}(t_{i}),\pmb{{r}}(t_{i}))}{\partial\pmb{u}(t_{i})} \,  \frac{d \pmb{{u}}(t_{i}))}{d\pmb{r}(t_{i})}+ \frac{\partial \pmb{G}(\pmb{{u}}(t_{i}),\pmb{{r}}(t_{i}))}{\partial\pmb{r}(t_{i})}. 
\end{align}
By using the recurrent quantum reservoir computer  ans\"atz, the Jacobian is  
\begin{multline}
\label{eq:10a}
{\mathbf{J}}(\pmb{r}(t_{i}), \pmb{{u}}(t_{i})) =  (1-\epsilon)\pmb{I} +   \epsilon \,\frac{\partial}{\partial \pmb{u}(t_{i}) } \langle  0|^{\otimes n} P^{\dag}(\pmb{r}(t_{i})) \, \,     \, \Xi^{\dag}(\pmb{u}(t_{i})) \,\pmb{\mathcal{W}} \, \Xi(\pmb{u}(t_{i})) \,  P(\pmb{r}(t_{i}))|0 \rangle ^{\otimes n} \, \pmb{W}_{out}^{{\textnormal T} } \\ + \epsilon \,\frac{\partial}{\partial \pmb{r}(t_{i}) } \langle 0|^{\otimes n} \, \,     \, P^{\dag}(\pmb{r}(t_{i})) \,\pmb{\mathcal{M}} \, P(\pmb{r}(t_{i})) \,  |0 \rangle^{\otimes n},
\end{multline}
where $\mathcal{M}^{(k)}=\Xi^{\dagger} \mathcal{W}^{(k)}\Xi$. This Jacobian can be calculated on quantum hardware with the parameter shift rule or linear addition of unitaries \citep{schuld2019evaluating}. In the RF-QRC (Section~\ref{sec:QRC}), $P(\pmb{r}(t_{i}))=\pmb{I}$, which simplifies the Jacobian to  
\begin{align}
\label{eq:10b}
{\mathbf{J}}(\pmb{r}(t_{i}), \pmb{{u}}(t_{i})) =  (1-\epsilon)\pmb{I} +   \epsilon \,\frac{\partial}{\partial \pmb{u}(t_{i}) } \langle  0|^{\otimes n} \, \,     \, \Xi^{\dag}(\pmb{u}(t_{i})) \,\pmb{\mathcal{W}} \, \Xi(\pmb{u}(t_{i})) \,  |0 \rangle ^{\otimes n} \, \pmb{W}_{out}^{{\textnormal T} }. 
\end{align}

As explained in section~\ref{sec:GS_CLE},  establishing whether GS occurs (or not) also requires the computation of the conditional Lyapunov exponents (CLEs), which are defined as the Lyapunov exponents of the Jacobian conditioned on the driving signal
\begin{align}\label{eq:5qrca}
\begin{split} 
{\mathbf{J}}(\pmb{r}(t_{i}))_{|_{\textnormal{CLE}}} \equiv  \frac{\partial {\pmb{G}}(\pmb{{r}}(t_{i}),\pmb{{u}}(t_{i}))}{\partial\pmb{r}(t_{i})}.
\end{split} 
\end{align}
Using the recurrent quantum reservoir computer ans\"atz yields the Jacobian for the computation of the CLEs
\begin{align}\label{eq:10}
\begin{split}
{\mathbf{J}}(\pmb{r}(t_{i}))_{|\textnormal{CLE}} &=  (1-\epsilon)\pmb{I} + \epsilon \,\frac{\partial}{\partial \pmb{r}(t_{i}) } \langle 0|^{\otimes n} \, \,     \, P^{\dag}(\pmb{r}(t_{i})) \,\pmb{\mathcal{M}} \, P(\pmb{r}(t_{i})) \,  |0 \rangle^{\otimes n}, 
\end{split}
\end{align}
which, in RF-QRCs, simplifies to  
\begin{align}\label{eq:C_RFQRC}
    {\mathbf{J}}(\pmb{r}(t_{i}))_{|\textnormal{CLE}} &=  (1-\epsilon)\pmb{I}. 
\end{align}}

Equation~\ref{eq:C_RFQRC} shows that the RF-QRC Jacobian is a perfectly conditioned and isotropic matrix with $N_r$  Lyapunov exponents equal to $1-\epsilon$. This is a key result of the paper: The Jacobian of the RF-QRC is constant (Eq.~\ref{eq:C_RFQRC}) and has negative CLEs. This mathematically shows that the RF-QRC fulfills, by design, condition (ii) in Table~\ref{tab: GS_eqv}, which means that the RF-QRC is, by design, asymptotically stable (hence $GS=ESP$). This is one the key results of this paper.

\section{Results}\label{sec:stab_qrc}

\subsection{Low-dimensional Lorenz-63 model}\label{sec:l63}

The first task is to forecast the chaotic dynamics and infer the invariant properties (Section~\ref{sec:stability}) for a three-dimensional Lorenz-63 system, which is a reduced-order model of a thermal convection flow \citep{75462} (Appendix~\ref{app:AppL96}). The time series data are obtained by the Runge-Kutta method. The numerical parameters are shown in Table~\ref{tab: param_RFQRC}. Each time series is divided into washout, training, and testing datasets \citep{brunton2022data}. The washout is the phase in which the transients are discarded to minimize the effect of initial conditions.  We perform a grid search to find good hyperparameters (Tikhonov regularization, $\beta$, and the leak rate, $\epsilon$) as shown in Table~\ref{tab: param_RFQRC}. For the Lorenz-63 system, we employ reservoirs with 7, 8, and 9 qubits. The results are shown for the reservoir size of 7 qubits. The parameters $\pmb{\alpha}$ of the random unitary ${V}(\pmb{\alpha})$ are sampled from a uniform distribution between $[0,4\pi]$. We train ten quantum reservoir networks, each with a different random seed, to reduce the effect of random initialization on the model performance. The ensemble average is used in the figures. The training length is 20 LTs, where LT is the Lyapunov time (Table~\ref{tab: param_RFQRC}). After training, in the autonomous phase, we deploy the RF-QRC to infer the invariant properties of the physical system.

\begin{table}[!h]
\centering
\caption{{Parameters for the tests on the Lorenz-63 and Lorenz-96 systems.}}
\label{tab: param_RFQRC}
\begin{tabular}{lllll}
\hline
Parameters &  & 3-D Lorenz-63  & 10-D Lorenz-96 & 20-D Lorenz-96 \\ \hline
Time step & $\Delta t$ & 0.01s & 0.01s & 0.01s \\
Leading Lyapunov exponent & $\lambda_{1}$ & 0.9056 &  1.2 & 1.5\\
Lyapunov Time & $LT$ & 1 LT = 110 steps & 1 LT = 83 steps & 1 LT = 66 steps\\
Washout steps & $N_{W}$ & 5 LT &  10 LT & 10 LT \\
Training steps & $N_{tr}$ & 20 LT & 200 LT  & 200 LT\\
Tikhonov regularization & $\beta$ & 1$\times10^{-9}$,1$\times10^{-12}$ & 1$\times10^{-9}$,1$\times10^{-12}$ & 1$\times10^{-9}$,1$\times10^{-12}$ \\
Input scaling & $\sigma$ & [0,1] & [0,1] & [0,1] \\ 
Leak rate & $\epsilon$ & 0.21 & 0.15 & 0.12 \\ 
Number of qubits & $n$ & 7 & 9 & 13 \\ 
Reservoir size & $N_{res}$ & 128 & 512 & 8192\\ 
Resevoir density & $D$ & Fully connected & Fully connected & Fully connected \\  \hline
\end{tabular}
\vspace*{-4pt}
\end{table}

\begin{table}[!h]
\centering
\caption{Lyapunov spectrum of the Lorenz-63 system. Comparison between the ground truth (target) and RF-QRC (inferred) model.}
\label{tab: lyap_L63}
\begin{tabular}{llll}
\hline
Exponents & Target & RF-QRC \\ 
\hline
1 & 0.9051 & 0.9173 \\
2 & 8.5$\times10^{-3}$ & 9.6$\times10^{-3}$  \\
3 & -14.56 & -14.65  \\ \hline
\end{tabular}
\vspace*{-4pt}
\end{table}

Both the ground truth and inferred Lyapunov spectra are obtained by autonomously evolving the network for 50 LTs. The quantum reservoir accurately infers the positive $\lambda_{1}$, neutral $\lambda_{2}$ and negative $\lambda_{3}$ Lyapunov exponents, as shown in Table~\ref{tab: lyap_L63}. 
Figure~\ref{fig:clv_stat} shows the distributions of the CLV  angles between the unstable-neutral subspaces, $\theta_{U,N}$,  the unstable-stable subspaces, $\theta_{U,S}$, and the neutral-stable subspaces, $\theta_{U,S}$. The  RF-QRC model accurately infers the long-term statistical distributions of the  CLV angles, which means that the RF-QRC can accurately infer the geometric structure of the tangent space.   Figure~\ref{fig:clv_angle} shows the angles between the subspaces on the chaotic attractor. When there are no tangencies, i.e., the subspaces are  separated, the system is hyperbolic with structurally stable dynamics and linear response \citep{lucarini2014mathematical}. 

\begin{figure}[!h]
\centering\includegraphics[width=6in]{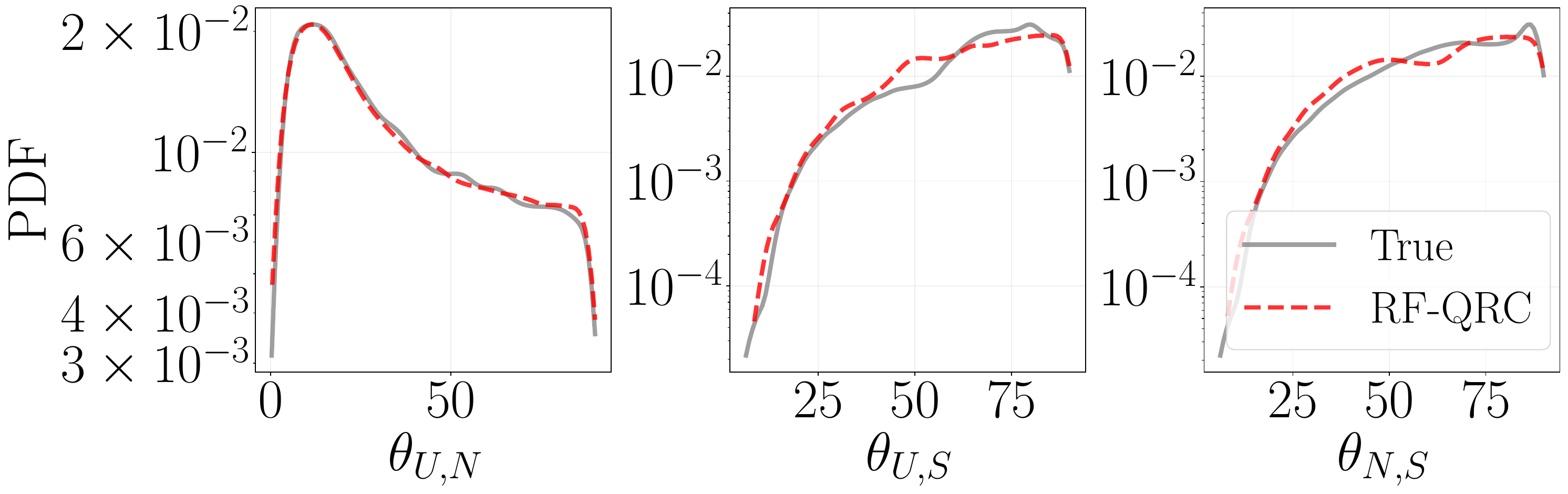}
\caption{Lorenz-63 system, RF-QRC with a reservoir of 7 qubits.  Probability density functions (PDFs) of the inferred the angles between the three subspace, where $U$ stands for unstable, $N$ for neutral, and $S$ for stable.}
\label{fig:clv_stat}
\end{figure}

\begin{figure}[!h]
\centering\includegraphics[width=6in]{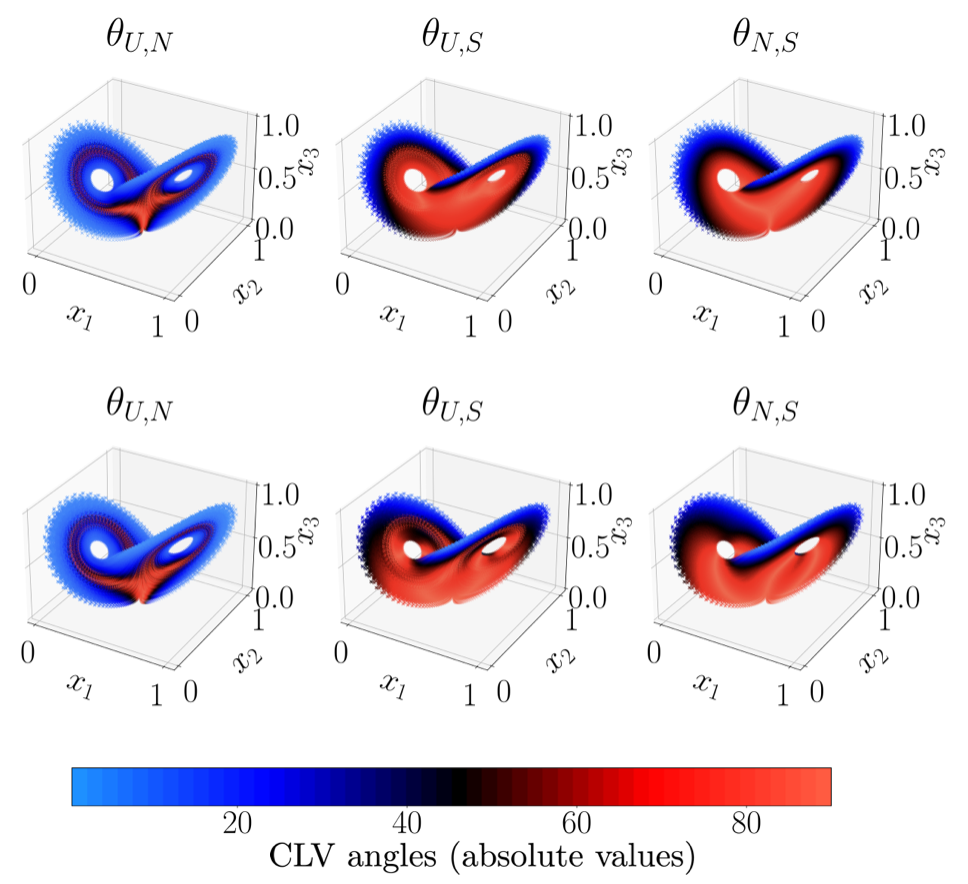}
\caption{Lorenz-63 system. Angles between subspaces. Top: Ground truth solution. Bottom: Prediction with the RF-QRC with a reservoir of 7 qubits. }
\label{fig:clv_angle}
\end{figure}

\subsection{Higher-dimensional Lorenz-96 model}\label{sec:l96}

We investigate the Lorenz-96 system \citep{75462} with 10 and 20 degrees of freedom (Appendix~\ref{app:AppL96}). Table~\ref{tab: param_RFQRC}  shows the model parameters.
We train ten different networks to reduce the effect of the random initialization due to $\pmb{\alpha}$. The training data is 200 LT long. After training with different reservoir sizes, we find that a good reservoir size is made of 9 qubits for a 10-dimensional case and of 13 qubits for a 20-dimensional Lorenz-96 system. In the prediction phase, the reservoir is evolved for 60 LTs. As shown in Figure~\ref{fig:l96_20D}, the RF-QRC accurately learns the Lyapunov spectrum from data. Table~\ref{tab: kydim} shows that the RF-QRC networks accurately infer the Kaplan-Yorke dimensions of the attractors. In conclusions, the RF-QRC can infer the physical invariant properties of the chaotic attractors in both low and higher dimensional systems. 

\begin{figure}[!h]
\centering\includegraphics[width=6in]{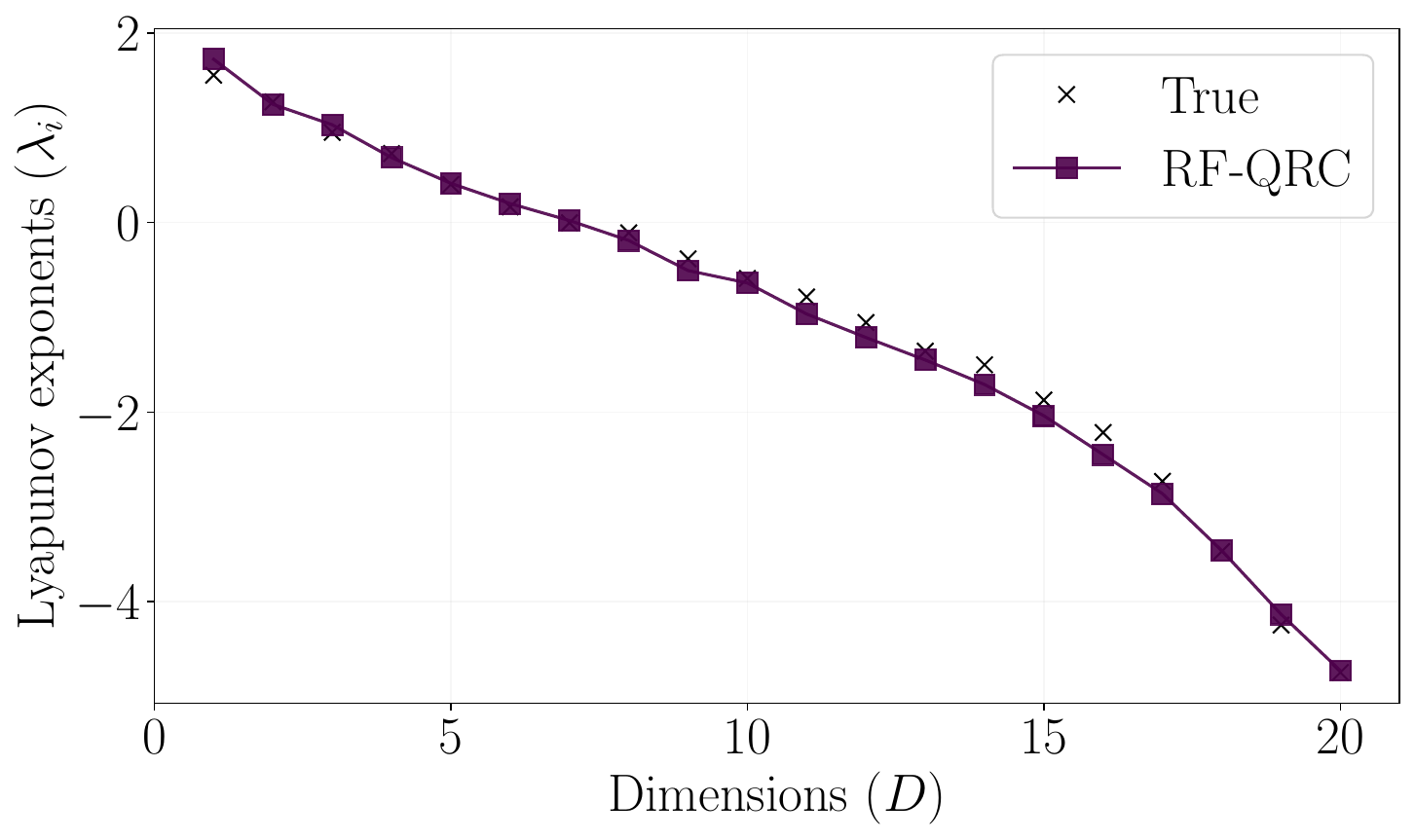}
\caption{Lyapunov spectrum of the Lorenz-96 system with 20 dimensions. The RF-QRC has a reservoir with 13 qubits.}
\label{fig:l96_20D}
\end{figure}

\begin{table}[!h]
\centering
\caption{Kaplan-Yorke dimensions. Comparison between the ground truth (target) and RF-QRC (inferred) model}
\label{tab: kydim}
\begin{tabular}{llll}
\hline
System & Target & RF-QRC & \% Error \\ 
\hline
Lorenz-63       &     2.06 & 2.06 & 0 \\
Lorenz-96 (10-D) &  6.52 & 6.58 & 0.92 \\
Lorenz-96 (20-D) &  13.40 &  13.25 & 1.1 \\ \hline
\end{tabular}
\vspace*{-4pt}
\end{table}

\section{Designing robust quantum reservoir computers}\label{sec:prac_qrc}

In Section~\ref{sec:stab_qrc}, we have shown that RF-QRCs can accurately forecast the chaotic dynamics whilst inferring the invariant physical properties of the physical system.  In this section, we provide practical guidelines for designing robust quantum reservoirs by exploiting the connection between quantum reservoir computers and GS made in Section~\ref{sec:QRC}. We compute the CLEs in the open-loop (training) phase for different leak rates, $\epsilon$, whilst keeping the quantum reservoir size fixed as in Section~\ref{sec:QRC}. The core idea is to analyse the conditional Lyapunov exponents  (Section~\ref{sec:stab_qrc}}) in the training phase to establish whether the quantum reservoir computers are robust (or not).  In the training phase, the Jacobian is provided by  Eq.~\ref{eq:5qrca}.  The pseudo-algorithm to compute the CLEs is shown in Algorithm~\ref{alg:CLE}.

\SetKwComment{Comment}{/* }{ */}
\begin{algorithm}[!h]
\caption{An algorithm to compute CLEs with RF-QRC}\label{alg:CLE}
{Choose $k$ leak rates $\epsilon$ in range [0,1]}\\
\For{each $\epsilon$}{

$\pmb{W} \gets random \in \mathbb{R}^{N_{r}\times D}$ \Comment*[r]{Initialize $D$ GSVs}
$\pmb{Q},\pmb{R} \gets QR{(\pmb{W})}$ \Comment*[r]{Orthonormalize GSVs}
$\pmb{W} \gets \pmb{Q} \in \mathbb{R}^{N_{r}\times D}$ \\
${N}_{QR} \gets N_{train}$ \Comment*[r]{Number of QR decompositions}
\textit{Save the time series of R for CLE  calculation} \\
Initialize $\pmb{\tilde{R}} \gets \pmb{0} \in \mathbb{R}^{D \times D \times N_{QR}}$ \\
$\pmb{J}_{CLE}$ = jacobian(RF-QRC) \Comment*[r]{Jacobian of the quantum circuit is computed in {Pennylane}\citep{bergholm2018pennylane} as in Eq.~\ref{eq:5qrca}}
\textit{Evolve the hidden state and GSVs simultaneously.}\\
\textit{Skip the initial transients for $N_{w}$ time steps}.\\
$n \gets 0$ \Comment*[r]{Increment the number of QR decompositions}
\For{$i = 0 : N_{tr}$}{
$\pmb{r}(t_{i+1}) = f(\pmb{r}(t_{i}))$  \Comment*[r]{QRC state update as in Eq.~\ref{eq:3c3}}
$\pmb{J} \gets \pmb{J}(\pmb{r}(t_{i}))$ \Comment*[r]{The updated Jacobian}
$\pmb{W} \gets \pmb{J}\pmb{W}$ \Comment*[r]{The variational equation}
$\pmb{Q},\pmb{R} \gets QR{(\pmb{W})}$ \Comment*[r]{QR at every time step}
$\pmb{W} \gets \pmb{Q}$ \\

  \If{$i > N_{w}$}{
    $\pmb{\Lambda}[:,n] \gets$ log(diag$[\pmb{R}])/dt$ \Comment*[r]{Save Finite time LEs} 
    $\pmb{\tilde{R}}[:,:,n] \gets \pmb{R}$ \Comment*[r]{Save \pmb{R}}
    $n = n+1$; \\
    \textbf{end}
  }
  \textbf{end}
}
$\lambda_{j} = \sum^{N_{QR}}_{i=0} \Lambda[j,i]/N_{train}$ \Comment*[r]{The \textit{j}th Conditional Lyapunov exponent}
$\lambda_{CLE} = \max(\lambda_{j})$ \Comment*[r]{The largest Conditional Lyapunov exponent}
}
\end{algorithm}

\begin{figure}
\centering\includegraphics[width=6in]{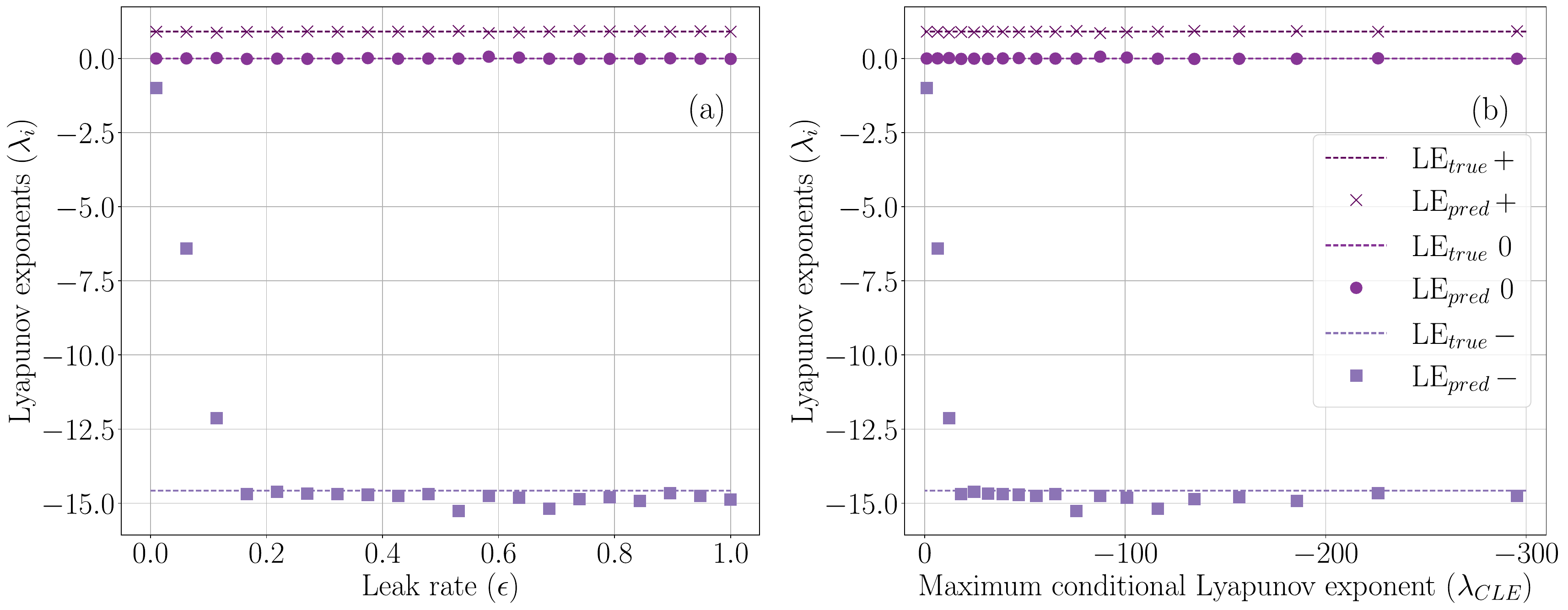}
\caption{Lorenz-63 in closed loop autonomous prediction phase.  (a) Lyapunov exponents vs leak rate. (a) Lyapunov exponents vs maximum conditional Lyapunov exponents.}
\label{fig:c4_LEs}
\end{figure}

First, Figure~\ref{fig:c4_LEs} (a) shows the  LEs inferred in closed loop for the Lorenz-63 system for $\epsilon$ in the range \{0,1\}. For $\epsilon$ $>$ 0.2, the spectrum is accurately inferred. For $\epsilon$ $<$ 0.2, only the positive and zero LEs are captured correctly. This is because, when the leak rate is small, the reservoir dynamics are slow, which means that the RF-QRC struggles to capture fast decaying perturbations associated with the negative LE. 

In Figure~\ref{fig:c4_LEs} (b), we show the maximum CLEs and LEs for the same $\epsilon$ as in Figure~\ref{fig:c4_LEs} (a). The magnitude of the maximum CLE grows with the leak rate $\epsilon$. For slower reservoir dynamics, when the maximum CLE is larger than most negative LE of the driving signal (max $\lambda_{CLE} > \lambda^{*}$, i.e., condition (ii) in Table~\ref{tab: CLE_GS}), the reservoir is unable to capture the negative LEs accurately. This leads to an incorrect increase in the Kaplan-Yorke dimension \citep{hart2024attractor}. Increasing the user-defined leak rate enables the tuning of the maximum CLE, thereby satisfying condition (iii) of Table~\ref{tab: CLE_GS}, which provides an accurate prediction of the entire spectrum. The same conclusion holds for the higher-dimensional Lorenz-96 system (Appendix~\ref{sec:appl96_qrc}).

\begin{figure}[!h]
\centering\includegraphics[width=5in]{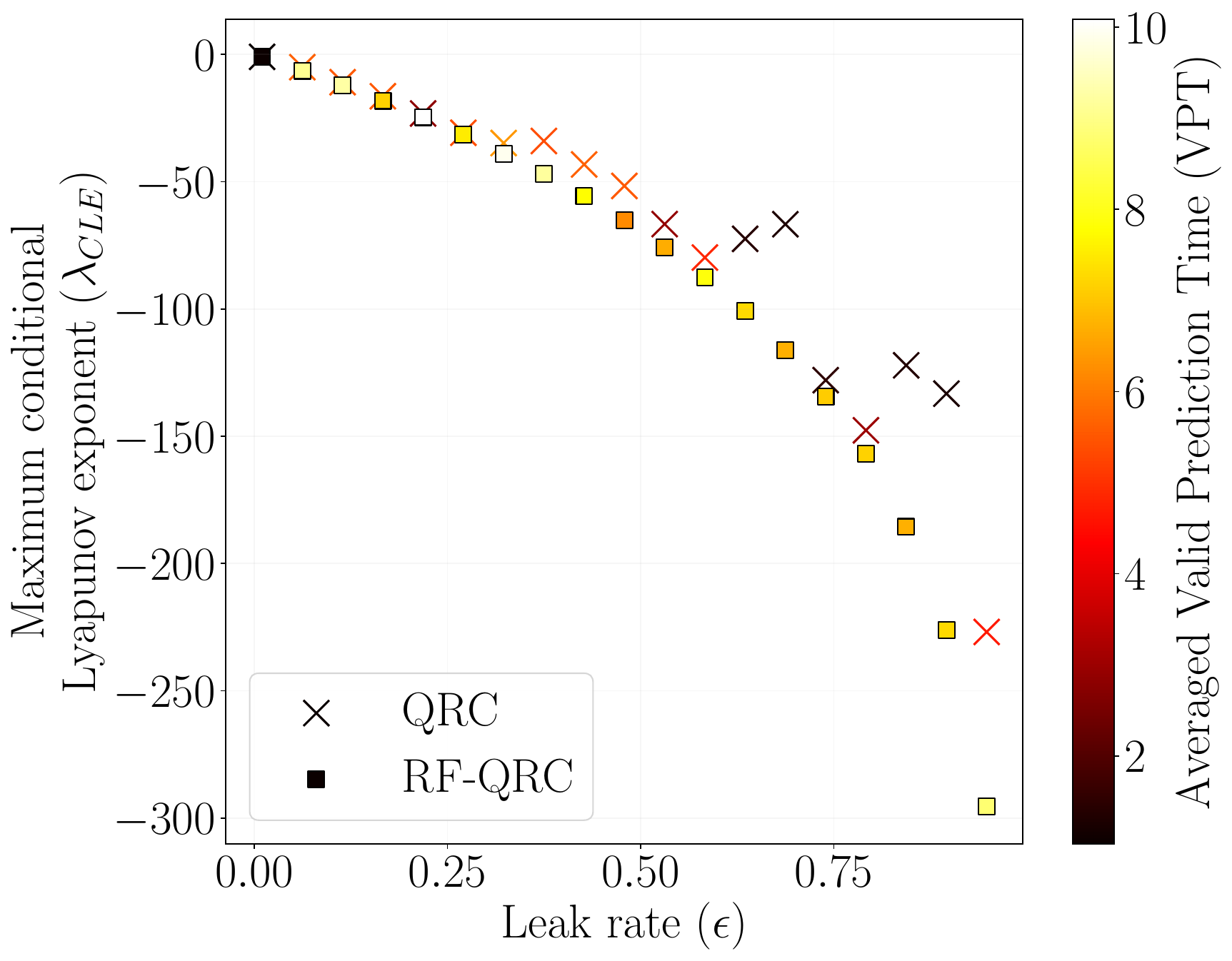}
\caption{Maximum conditional Lyapunov exponents of quantum reservoir computers with recurrence (QRCs) and recurrence-free (RF-QRCs). Each point value is the ensemble mean over ten autonomous predictions from different initial points.}
\label{fig:LCV_comb}
\end{figure}

Finally, in Figure~\ref{fig:LCV_comb} we show the maximum CLEs as functions of the leak rate for both recurrent and recurrence-free quantum reservoir computers (QRCs and RF-QRCs, respectively). The conditional Lyapunov exponents (CLEs) remain negative for all leak rates, which means that the reservoir dynamics are stable. The colour of each point in Figure~\ref{fig:LCV_comb} quantifies the short-term prediction of the reservoirs with the valid prediction time \citep{vlachas_backpropagation_2020}. The results in Figure~\ref{fig:LCV_comb} show that the RF-QRC performance is robust and predictable over a larger range of leak rates than that of QRCs. In detail, in RF-QRCs, there exists an injective (monotonic) relationship between the leak rate and maximum CLE. The network approaches the behaviour of an Extreme Learning Machine (ELM) \citep{xiong2023fundamental} when $\epsilon=1$, with the magnitude of maximum CLE approaching infinity (see Eqs. \ref{eq:C_RFQRC} and \ref{f4jw0f582j08f24}). A very large magnitude of the maximum CLE makes the reservoir unstable and sensitive to noise. On the other hand, in  QRCs, the relationship between $\epsilon$ and the maximum CLE is not injective (not monotonic). QRCs are stable for intermediate $\epsilon$ values ($0.25 < \epsilon < 0.6$), whereas the RF-QRC is stable for a larger range, $\epsilon > 0.2$. In these regions, QRCs and RF-QRCs can be used for both short-term predictions and inference of invariant properties from chaotic time series.
In Appendix~\ref{app:stab_recurrent}, we show the results obtained with the  QRCs, which further corroborate the robustness of the RF-QRC. 

\subsection{Influence of noise}\label{sec:noise}

In this section, we analyse the effect of different types of noise on the performance of RF-QRCs.  We analyse the role of finite sampling noise, which arises from the probabilistic nature of quantum mechanics, and, therefore, it is naturally present in Fault-Tolerant Quantum Computers (FTQCs) \citep{preskill1998fault}. Additionally, we analyse types of noise that promote dissipation (loss of information) in the reservoir update. We analyse incoherent errors due to the interaction of the quantum hardware with its environment, i.e., depolarizing noise and amplitude damping channels \citep{9781107002173}. Previous works have proposed  dissipation from noise as a resource to improve quantum reservoir computers \citep{sannia_dissipation_2022,fry2023optimizing,domingo_taking_2023}. We show that tuning noise intensity helps satisfy $GS=ESP$.  

\begin{table}[!h]
\centering
\caption{Types of noise analysed.}
\label{tab: noiseparam}
\begin{tabular}{llll}
\hline
Cause & Type of noise & Parameters & Number of qubits \\ 
\hline
Finite sampling        &  Projective measurement  &  $S$ = \{1000,5000,10000, &  $n$ = 8 \\ 
 & & \quad \quad \quad \quad 25000,50000\} &\\
Depolarizing channel &  Unital, incoherent      &  $p$ = \{0.001,0.01,0.05,0.1\}  & $n$ = 7 \\
Amplitude damping    &  Non-unital, incoherent  &  $p$ = \{0.001,0.01,0.05,0.1\}  & $n$ = 7 \\ \hline
\end{tabular}
\vspace*{-4pt}
\end{table}

\subsubsection{Sampling noise}

To recover the classical information on the time series, the quantum reservoir state update (Eq.~\ref{eq:3c3}) needs to be measured in the computational basis (Eq.~\ref{eq:3c4}). 
Mathematically, the measurement makes the quantum wave function collapse randomly on one of the eigenbases; therefore, multiple measurements (also known as shots $S$) must be performed to obtain a statistical estimate of the reservoir state. 
The effect of finite sampling noise \citep{mujal_time_2023,kobayashi2024feedback,ahmed2025optimal,xiong2023fundamental,hu2023tackling} on the reservoir state can be modelled as an additive stochastic term $\boldsymbol{\zeta}(t)$ with zero mean
%
\begin{align}\label{eq:4a2} 
\begin{split} 
{{\pmb{{r}}}}(t)  = \bar{\pmb{{r}}}(t) + \frac{1}{\sqrt{S}} \boldsymbol{\zeta}(t).
\end{split}
\end{align}
In the limit of infinite measurements, $S\rightarrow \infty$, the noisy estimator ${\pmb{{r}}}(t)$ converges to the noise-free reservoir signal $\bar{\pmb{r}}$.
 The goal of this section is to analyse the effect that sampling noise has on the conclusions drawn in the noise-free scenarios of Section~\ref{sec:stab_qrc}. We evaluate the performance of inferring the Lyapunov exponents for different shots, $S$ (Table~\ref{tab: noiseparam}).

\begin{figure}[!h]
\centering\includegraphics[width=6in]{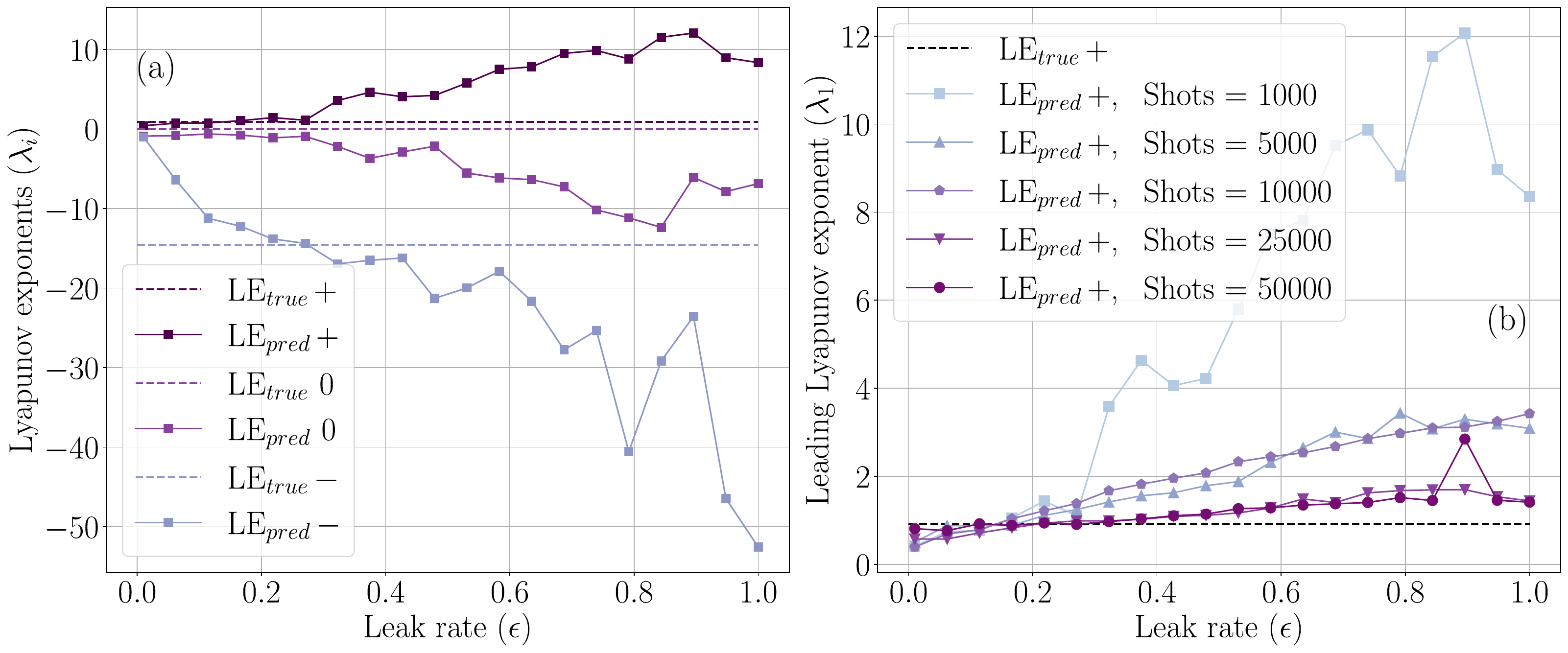}
\caption{Effect of sampling noise on the RF-QRC performance with 8 qubits. (a) Lyapunov spectrum for 1000 shots. (b) Leading Lyapunov exponents for different shots. }
\label{fig:ShotsLEs}
\end{figure}

Figure~\ref{fig:ShotsLEs} (a)  shows the results of a RF-QRC with an 8-qubit reservoir sampled with 1000 shots at each time step during training and prediction. The results are the outcome of averaging the performances over five different random seeds for the $V(\pmb{\alpha})$ unitary. Even for a number of finite samples, RF-QRC is able to accurately infer the LEs of the Lorenz-63 system when the leak rate is small (the memory is high). This means that in the presence of noise, the reservoir dynamics must be slower to infer the LEs (noise promotes dissipation, thereby requiring larger memory). However, reducing the leak rate to values close to zero, or operating near the edge of chaos ($\lambda_{CLE}=0$ in Table~\ref{tab: CLE_GS}), results in a loss of generalized synchronization. In this case, in the presence of noise due to finite sampling, the hyperparameter leak rate should be tuned in the range $\{0.2,0.3\}$. Even for finite samples (e.g., 1000 in this case), there exists a good range of leak rates for which the RF-QRC infers the stability properties accurately. Finding this optimum range can vary depending on the analysed chaotic system and the sample size. In Figure~\ref{fig:ShotsLEs} (b), we perform a parametric study with the number of shots (Table~\ref{tab: noiseparam}). The noise intensity influences the performance of the RF-QRC and the range of good hyperparameters $\epsilon$. As expected, by increasing the number of shots, the accuracy  increases. 

\subsubsection{Incoherent noise}

\begin{figure}[!h]
\centering\includegraphics[width=6in]{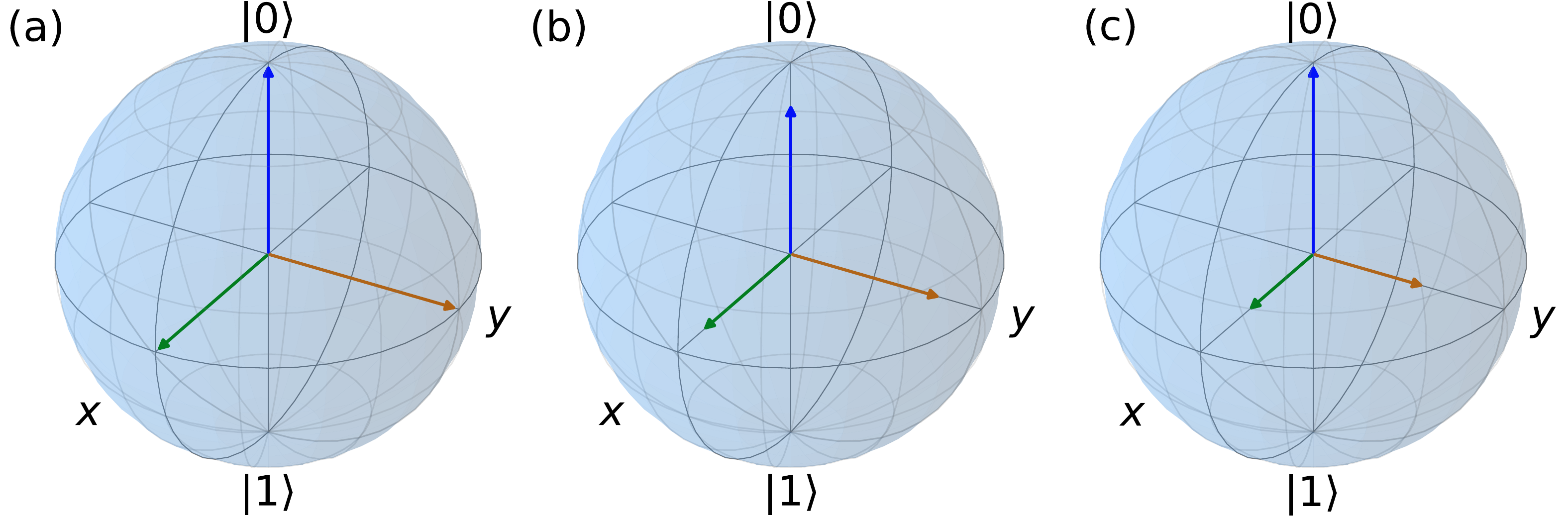}
\caption{Bloch spheres of the quantum state of a qubit for (a) noiseless unitary evolution, (b) uniform contraction of the state due to depolarizing noise, and (c) non-uniform contraction of the state due to amplitude damping noise.}
\label{fig:bloch}
\end{figure}

A quantum hardware state, represented by a density operator  $\rho_{H}$, is entangled with the environment, $\rho_{env}$, to form a combined system-environment state, which is modelled as a product state $\rho_{H} \otimes \rho_{env}$. To consider the effect of a unitary  transformation, $\mathcal{U}$, acting on the quantum hardware, we take a partial trace over the environment to obtain the reduced state of the system \citep{9781107002173}
\begin{align}\label{eq:n0}
\begin{split} 
    \mathcal{N}(\rho_{H}) = \textnormal{tr}_{env} [\mathcal{U} (\rho_{H} \otimes \rho_{env}) \mathcal{U}^{\dag}],   
\end{split}
\end{align}
where $\dagger$ is the conjugate transpose, and   $\mathcal{N}$ is a noisy channel, which takes the quantum hardware state $\rho_{H} \rightarrow \mathcal{N}(\rho_{H})$. Let $|e_{m}\rangle$ be an orthonormal basis for the (finite-dimensional) state space of the environment, and $\rho_{env} = |e_{0}\rangle \langle e_{0}|$ be the initial state of the environment (assuming pure state). Eq.~\ref{eq:n0} in \textit{operator-sum} representation can be written as 
\begin{align}\label{eq:n01}
\mathcal{N}(\rho_{H}) &= \sum_{m} \langle e_{m}| \, \mathcal{U} \, \, [(\rho_{H} \otimes |e_{0}\rangle \langle e_{0}|] \, \,
\mathcal{U}^{\dag} |e_{m}\rangle,   \\
 &= \sum_{m} K_{m} \, \rho K_{m}^{\dag}, 
\end{align}
where $K_{m} \equiv \langle e_{m}| \mathcal{U}|e_{0}\rangle$ is an operator on the state space of the quantum system. This representation is also known as the Kraus representation, in which $K_{m}$ are the Kraus operators, which fulfill $\sum_{m} K_{m} K_{m}^{\dag} = I $  \citep{9781107002173}. The maximum number of Kraus operators to model incoherent noise and, consequently, the size of the noisy state, scales quadratically with the size of the Hilbert space $(2^{n})^{2}$.
We consider the effect of amplitude damping and depolarizing channels (Figure~\ref{fig:bloch}). These types of noise have been previously studied in quantum natural reservoirs  \citep{suzuki_natural_2022,kubota_temporal_2023}, using dissipation as a resource \citep{sannia_dissipation_2022} and with artificial noise channels with tunable noise parameters \citep{fry2023optimizing}. 
We consider the  density operator, $\rho_{H}$, that represents the quantum state of one qubit
\begin{align}\label{eq:n2}
\begin{split}
    \rho_{H} = 
\begin{pmatrix}
\rho_{00} & \rho_{01} \\
\rho_{01}^{\dag} & \rho_{11}
\end{pmatrix},
\end{split}
\end{align}
 The amplitude damping channel applied to this state can be modelled with the Kraus operators $K_{0}$ and $K_{1}$ 
\begin{align}\label{eq:n3}
\begin{split}
K_{0} = 
\begin{pmatrix}
1 & 0 \\
0 & \sqrt{1-p}
\end{pmatrix}
,
   K_{1} = 
\begin{pmatrix}
0 & \sqrt{p} \\
0 & 0
\end{pmatrix},
\end{split}
\end{align}
where $0 \leq p \leq 1 $ is the noise intensity. The Kraus operators can be combined with Eq.~\ref{eq:n01} to build the density matrix $\mathcal{N}_{AD}(\rho_{H})$ \citep{9781107002173}. Physically, this type of noise represents the asymmetric shrinking of the Bloch sphere, thus, it is referred to as non-unital noise  (Figure~\ref{fig:bloch} c). We vary the intensity of the noise channel with  $p$, as shown in Table~\ref{tab: noiseparam}, for a 7-qubit quantum reservoir system. The visualization of the noise channel in the quantum circuit is shown in Figure~\ref{fig:noise_cirq} with $\eta$ being the noise channel. 

\begin{figure}[!h]
\centering\includegraphics[width=6in]{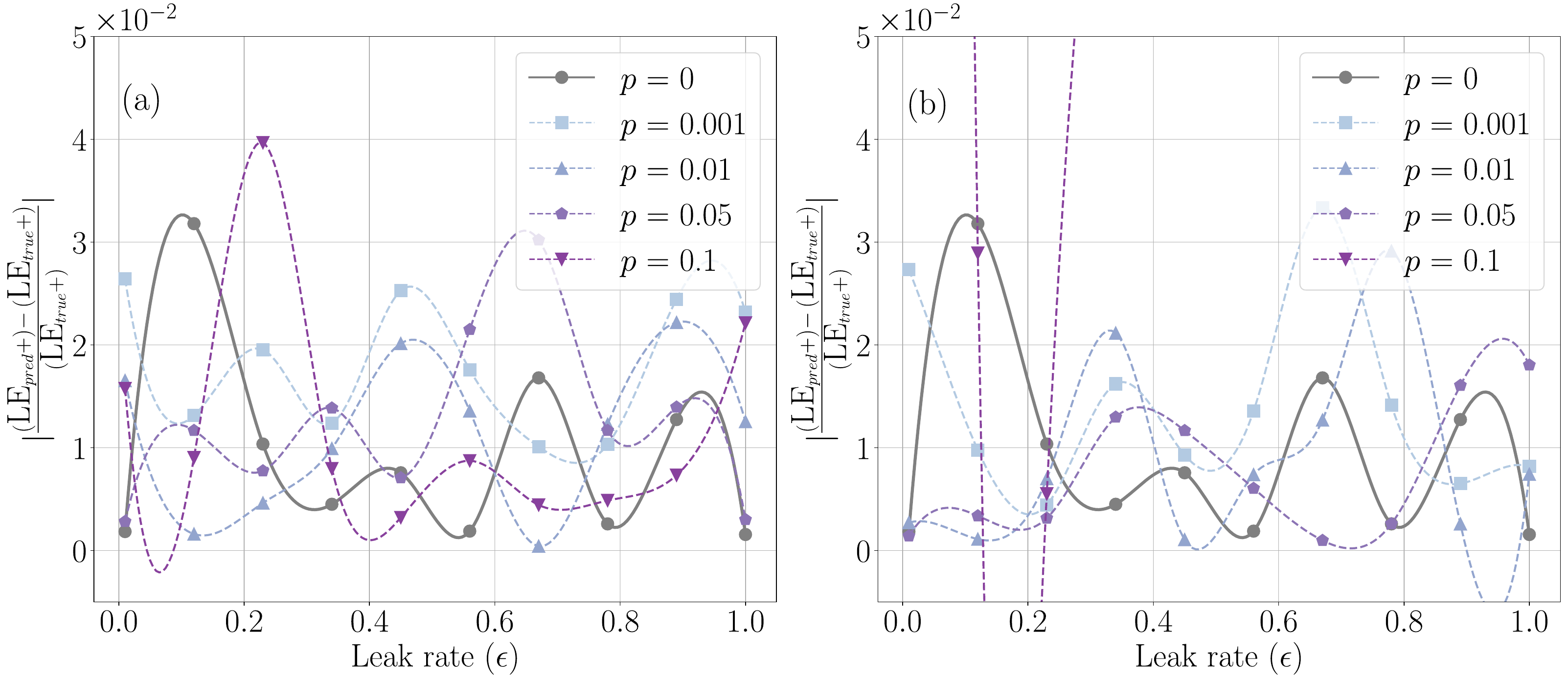}
\caption{Error on the leading Lyapunov exponent with a RF-QRC with 7 qubits. Each result is the ensemble average over 3 different realizations of random seeds.  (a) Effect of amplitude damping noise.  (b) Effect of depolarizing noise.}
\label{fig:ADLEs_all}
\end{figure}

The second type of noise analysed  is the depolarizing channel, which for a single qubit can be represented by \footnote{The associated Kraus representation could also be derived with 4 Kraus operator $\{K_{0}, K_{1}, K_{2}, K_{3}\}$ and corresponding Pauli matrices $\{I, X, Y, Z\}$ as shown in \citep{preskill1998lecture}.}
\begin{align}\label{eq:n4}
\begin{split}    
    \mathcal{N}_{DP}(\rho_{H}) = (1-p) \, \rho_{H} + p \, \frac{I}{2}.    
\end{split}
\end{align}

The depolarizing noise shrinks the Bloch sphere uniformly, i.e., it is a unital type of incoherent noise (Figure~\ref{fig:bloch} b). With $p = 1$, the channel returns the maximally mixed state for any input state $\rho_{H}$, which corresponds to the complete contraction of the Bloch sphere to a single point given by $\frac{I}{2} $. We vary the intensity of the noise channel (Table~\ref{tab: noiseparam} and Figure~\ref {fig:noise_cirq}).
First, Figure~\ref{fig:ADLEs_all} (a) shows the impact of the amplitude-damping noise channels on the RF-QRC performance. We show the error between the inferred and target leading Lyapunov exponents ($\lambda_{1}$) for different noise intensities and leak rates. Each data point in Figure~\ref{fig:ADLEs_all} is computed by ensemble averaging over three random seeds of the $V(\pmb{\alpha})$ unitary. The case with $p = 0 $ corresponds to the ideal noise-free state-vector emulation. We increase the amplitude damping noise intensity from $0.001 - 0.1$     ($0.1-10\%$ noise intensity). The largest error in the prediction on $\lambda_{1}$  is $\approx0.04$, which occurs for $p = 0.1$. For other noise intensities, when the reservoir dynamics are slower (small leak rate), the prediction performance improves over the noise-free scenario. This shows that adding amplitude damping noise promotes dissipation,  thereby strengthening GS. The addition of noise in the faster reservoir dynamics region (i.e., large leak rate) does not improve the performance beyond the ideal noise-free case.
Figure~\ref{fig:ADLEs_all} (b)  shows the results on the depolarizing noise channel. For  $p = 0.1$, the reservoir cannot infer the correct value of $\lambda_{1}$, and the maximum relative error is  $\approx0.2$. As in amplitude damping, when the reservoir dynamics are slower (i.e., small leak rate), the performance improves with the addition of depolarizing noise channels, i.e., noise makes the RF-QRC more robust and accurate than the ideal noise-free case. In conclusion, noise adds dissipation, which, in turn, promotes GS (and consequently ESP). This can be from Eq.~\ref{eq:3d2}, in which noise ensures a contractive map  ($\tilde{\gamma} < 1$).

\section{Conclusions}\label{sec:conclusion}
We propose criteria and methods for the analysis and design of robust quantum reservoir computers for chaotic time series forecasting. 
Key to the criteria and methods is 
 interpreting quantum reservoir computers as coupled dynamical systems, which can be analysed with generalized synchronization (GS) from dynamical systems theory. The core element of the framework is the Jacobian, which is analytically derived for both recurrent quantum reservoir computers (QRCs) and recurrence-free quantum reservoir computers (RF-QRCs).
First, we show that quantum reservoir computers can accurately predict the chaotic dynamics and their scalar invariant properties, such as Lyapunov spectra, attractor dimensions, and their geometric invariant properties, such as the covariant Lyapunov vectors. 
 We  test the framework on low and higher-dimensional chaotic systems (Lorenz-63 and Lorenz-96 systems). We show that both QRCs and RF-QRCs can infer correctly the chaotic dynamics, their long-term statistics, and the invariant properties. The RF-QRCs are more accurate and robust across larger ranges of hyperparameters.  
 Second, we propose a criterion for quantum reservoir computers to fulfill the echo state property (ESP), which is a necessary condition for the robust design of reservoir computers. The shorthand for the criterion is $GS=ESP$, which means that if the quantum reservoir computer is in generalized synchronization with the training data, then it fulfills the echo state property, and vice versa. 
Third, we provide a method to design robust quantum reservoir computers and evaluate when $GS=ESP$ holds. The method is based on the comparison between the Lyapunov exponents and the conditional Lyapunov exponents (CLEs) of a quantum reservoir computer.  We show that recurrence-free quantum reservoir computers (RF-QRCs) are conditionally stable, which means that they satisfy $GS=ESP$, by design,  across a larger range of hyperparameters than that of recurrent quantum computers. 
Finally, we consider noise caused by finite sampling, depolarizing channels, and amplitude damping. We show that the dissipation caused enhances the contraction rate of the  quantum reservoir computer map. This shows that noise can be exploited to make QRCs and RF-QRCs more robust.   
RF-QRCs accurately infer the chaotic dynamics and its invariant properties even in noisy scenarios. We find larger sets of hyperparameters in which noise improves the accuracy of RF-QRCs with respect to the noise-free setting. 
This work opens opportunities for employing quantum reservoir computers for chaotic time series forecasting, and designing robust quantum reservoirs for implementation on near-term quantum devices.


\paragraph{Acknowledgements}{O.A. thanks the Department of Aeronautics EPSRC studentship for funding the PhD. The authors acknowledge financial support from the UKRI New Horizon grant EP/X017249/1. L.M. is grateful for the support from the ERC Starting Grant PhyCo 949388, F.T. acknowledges support from the UKRI AI for Net Zero grant EP/Y005619/1. O.A. thanks Elise \"Ozalp for discussions on stability analysis. }

\bibliographystyle{IEEEtran}
\bibliography{bibliography} 

\begin{appendix}

\section{The computation of LEs and CLVs}\label{app:algo}

We use the procedure to compute LEs and CLVs as described in \citep{margazoglou_stability_2023}. Here, we outline the algorithm for computing these stability properties on classical computers using a quantum reservoir as a feature generator. Gram-Schmidt vectors (GSVs) are randomly initialized to span a linearly independent orthonormal basis.
After computing the LEs and saving  ${\pmb{\tilde{R}}}$ and ${\pmb{\tilde{Q}}}$, both of these matrices can be used to compute the covariant Lyapunov vectors. The process of computing CLVs, after generating ${\pmb{\tilde{R}}}$ and ${\pmb{\tilde{Q}}}$ with a quantum reservoir, uses Algorithm 2 of  \citep{margazoglou_stability_2023}.
\SetKwComment{Comment}{/* }{ */}

\begin{algorithm}[!h]
\caption{An algorithm to compute LEs with RF-QRC}\label{alg:two}
$\pmb{W} \gets random \in \mathbb{R}^{N_{r}\times D}$ \Comment*[r]{Initialize $D$ GSVs}
$\pmb{Q},\pmb{R} \gets QR{(\pmb{W})}$ \Comment*[r]{Orthonormalize GSVs}
$\pmb{W} \gets \pmb{Q} \in \mathbb{R}^{N_{r}\times D}$ \\
${N}_{QR} \gets N_{test}$ \Comment*[r]{Number of QR decompositions}
\textit{Save the time series of R and Q for CLVs calculation} \\
Initialize $\pmb{\tilde{R}} \gets \pmb{0} \in \mathbb{R}^{D \times D \times N_{QR}}$ \\
Initialize $\pmb{\tilde{Q}} \gets \pmb{0} \in \mathbb{R}^{N_{r} \times D \times N_{QR}}$ \\
Initialize $\pmb{{\Lambda}} \gets \pmb{0} \in \mathbb{R}^{D \times N_{QR}}$ \\
$\pmb{J}$ = jacobian(QRC) \Comment*[r]{Jacobian of the quantum circuit is computed in {Pennylane}\citep{bergholm2018pennylane} as in Eq.~\ref{eq:10}}
\textit{Evolve the hidden state and GSVs simultaneously.}\\
\textit{Skip the initial transients for $N_{w}$ time steps}.\\
$n \gets 0$ \Comment*[r]{Increment the number of QR decompositions}
\For{$i = 0 : N_{test}$}{
$\pmb{r}(t_{i+1}) = f(\pmb{r}(t_{i}))$  \Comment*[r]{QRC state update as in Eq.~\ref{eq:3c3}}
$\pmb{u}(t_{i+1})=[\pmb{r}(t_{i+1})]^{{\textnormal T}}\pmb{W}_{out}$ \\
$\pmb{J} \gets \pmb{J}(\pmb{r}(t_{i}))$ \Comment*[r]{The updated Jacobian}
$\pmb{W} \gets \pmb{J}\pmb{W}$ \Comment*[r]{The variational equation}
$\pmb{Q},\pmb{R} \gets QR{(\pmb{W})}$ \Comment*[r]{QR at every time step}
$\pmb{W} \gets \pmb{Q}$ \\

  \If{$i > N_{w}$}{
    $\pmb{\Lambda}[:,n] \gets$ log(diag$[\pmb{R}])/dt$ \Comment*[r]{Save Finite time LEs} 
    $\pmb{\tilde{R}}[:,:,n] \gets \pmb{R}$ \Comment*[r]{Save \pmb{R}}
    $\pmb{\tilde{Q}}[:,:,n] \gets \pmb{Q}$ \Comment*[r]{Save \pmb{Q}}
    $n = n+1$; \\
    \textbf{end}
  }
  \textbf{end}
}
$\lambda_{j} = \sum^{N_{QR}}_{i=0} \Lambda[j,i]/N_{test}$ \Comment*[r]{The \textit{j}th Lyapunov exponent}
\end{algorithm}

\section{Physical systems}
\label{app:AppL96}

The Lorenz-63 system \citep{75462} is a reduced-order model of thermal convection flow governed by 
\begin{align}\label{eq:3A}
\begin{split}
    \frac{dx_{1}}{dt} = \sigma \, (x_{2}-x_{1})
\end{split}
\end{align}
\begin{align}\label{eq:3B}
\begin{split}
    \frac{dx_{2}}{dt} = x_{1} \, (\rho-x_{3}) - x_{2}, 
\end{split}
\end{align}
\begin{align}\label{eq:3C}
\begin{split}
    \frac{dx_{3}}{dt} =  x_{1}x_{2}-\beta{x_{3}},
\end{split}
\end{align}
where $\sigma$, $\rho$, $\beta$ are system parameters. We take [$\sigma$ , $\rho$ , $\beta$] = [10, 28, 8/3] to ensure chaotic behavior of the system. 
The Lorenz-96 model \citep{75462} is a system of coupled ordinary differential equations that describes the large-scale behavior of the mid-latitude atmosphere, and the transfer of a scalar atmospheric quantity, governed by 

\begin{align}\label{eq:18}
    \frac{dx_{i}}{dt} =  (x_{i+1}-x_{i-2}) \, x_{i-1}-{x_{i}} + F  , \quad \quad \quad i = 1, \dotsc ,m
\end{align}
where $F$ is the external body forcing term which we set to $F = 8$ to achieve chaotic behavior. We apply periodic boundary conditions, i.e. $x_{1} = x_{m+1}$, and study the reduced order model of Lorenz-96 with ten and twenty dimensions corresponding to $m=10$ and $m=20$. 

\section{Stability analysis of a 10-dimensional Lorenz-96 model}\label{sec:appl96_qrc}

\begin{figure}[!h]
\centering
\includegraphics[width=6in]{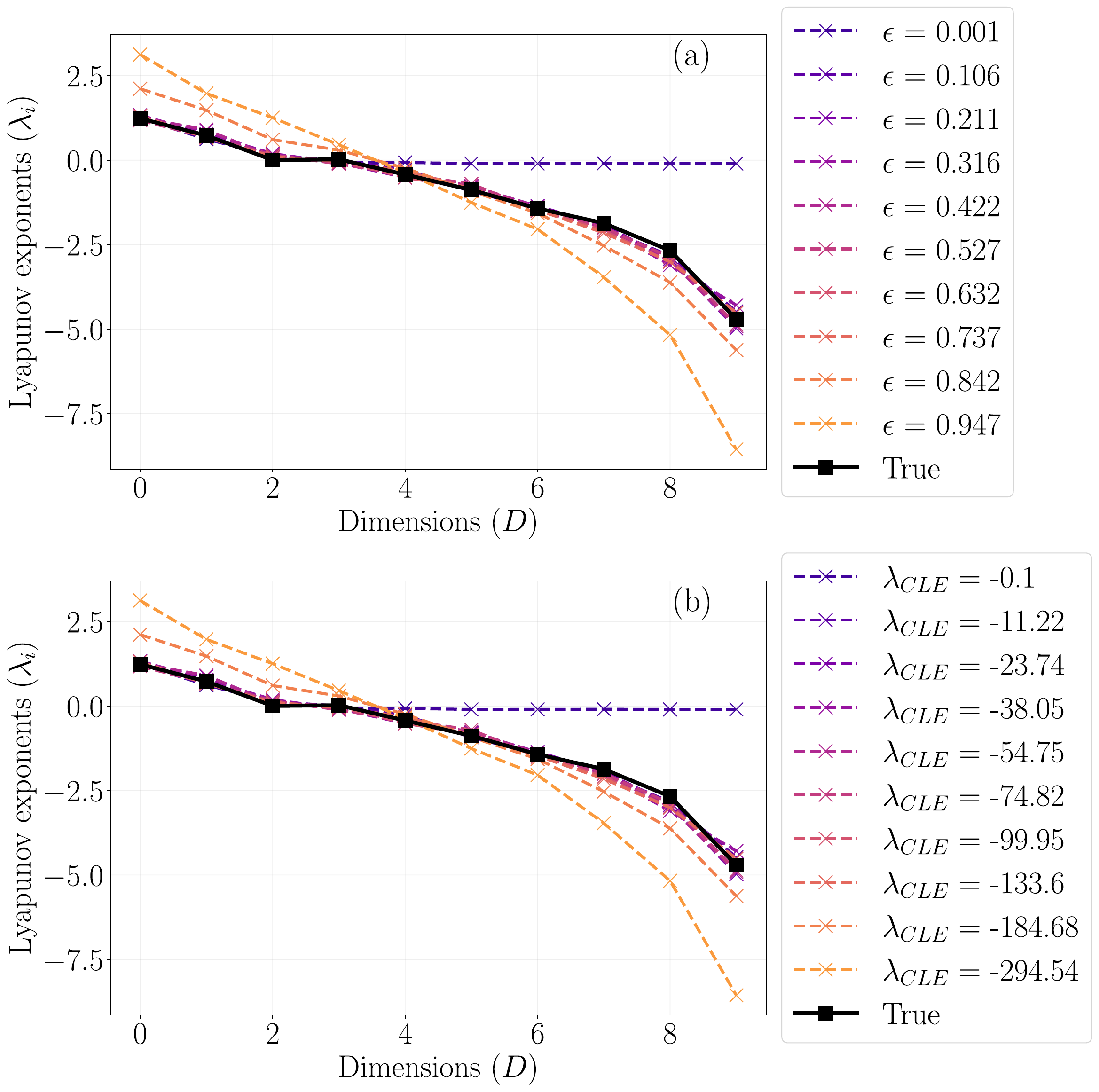}
\caption{Lyapunov spectrum of Lorenz-96 systems with 10 dimensions. Comparison between ground truth and RF-QRC-inferred Lyapunov spectrum for a reservoir size of 10 qubits. (a) Various leak rates (b) Various maximum $\lambda_{CLE}$ corresponding to leak rate in (a)}
\label{fig:l96_10D_comb}
\end{figure}

We extend the analysis about designing practical quantum reservoirs, shown in Section~\ref{sec:prac_qrc} for the Lorenz-63 system, to the 10-dimensional Lorenz-96 system. Specifically, we analyse the influence of the leak rate hyperparameter $\epsilon$ and the magnitude of the maximum conditional Lyapunov exponent  on the ability of the reservoir to infer Lyapunov exponents on the unseen test data set. The training procedure is the same as described in Section~\ref{sec:l96} and other model hyperparameters are shown in Table~\ref{tab: param_RFQRC}. 
In Figure~\ref{fig:l96_10D_comb}a, we vary the leak rate from $\epsilon=0.001$ to $\epsilon = 1.0$ and compute the inferred Lyapunov exponents of the model. In Figure~\ref{fig:l96_10D_comb} b, the colour bar indicates the maximum $\lambda_{CLE}$ value associated with the corresponding leak rate in Figure~\ref{fig:l96_10D_comb} a. These results are in line with the findings of the previous analysis from Section~\ref{sec:prac_qrc}. The  maximum CLEs and their effects on the inference task are shown in Figure~\ref{fig:l96_10D_comb} b. When the reservoir dynamics are faster and the leak rate is closer to 1, the reservoir over-predicts the positive and under-predicts the negative Lyapunov exponents. Because the reservoir dynamics are faster, the reservoir is also very sensitive to noise in this region. Therefore, depending on the learning tasks and form of noise, the rate of reservoir dynamics must be carefully tuned with the hyperparameter leak rate. 

\section{Stability in recurrent quantum reservoir computers (QRCs)}\label{app:stab_recurrent}

\begin{figure}[!h]
\centering
\includegraphics[width=6in]{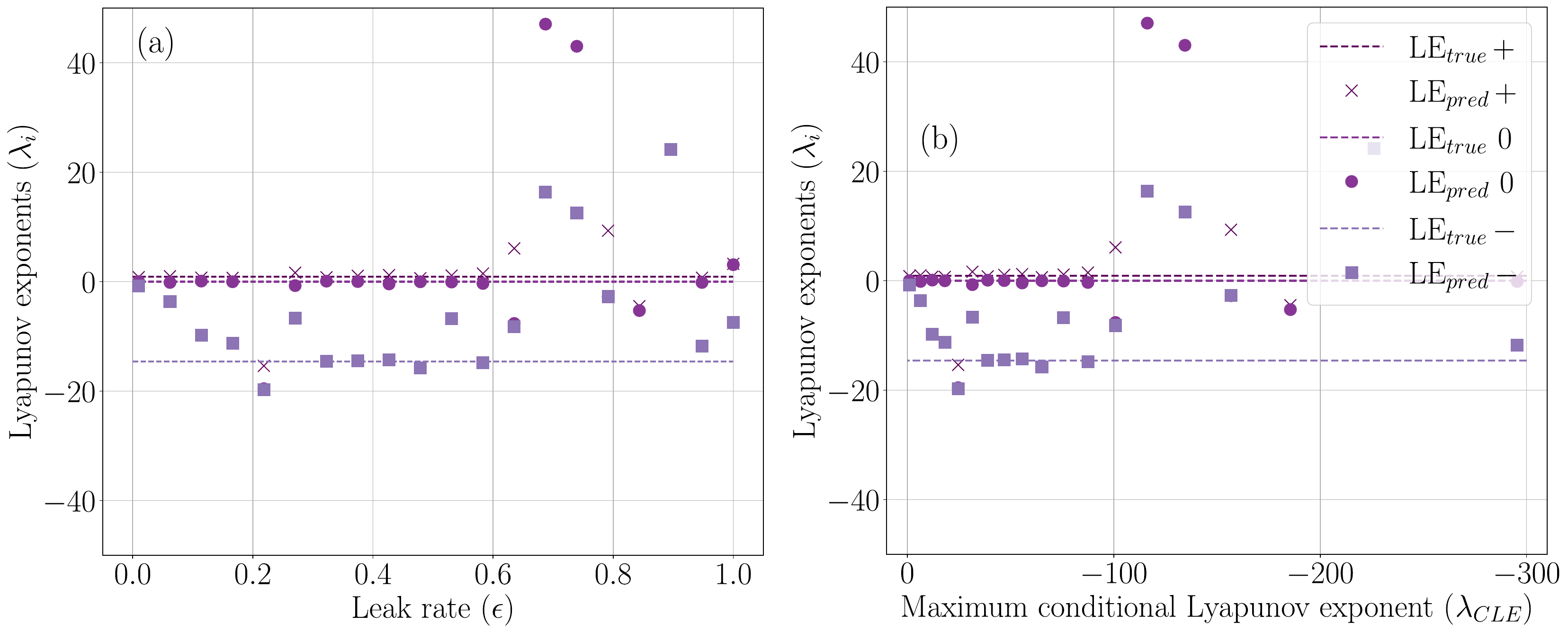}
\caption{Lorenz-63 system autonomous prediction of Lyapunov spectrum with QRCs (a) Dynamical system's Lyapunov exponents vs Maximum conditional Lyapunov exponents of response (b) Dynamical system's Lyapunov exponents vs Leak rate. In both plots, three points on each vertical line correspond to the three inferred Lyapunov exponents. For very small leak rates, reservoir dynamics are slower, and negative LE is not inferred correctly. QRC  performance is optimum for intermediate $\epsilon$ between \{0.35,0.5\} but performs inaccurately  with $\epsilon > 0.6$ (as opposed to RF-QRC in Figure~\ref{fig:c4_LEs}).}
    \label{fig:c2_LEs}
\end{figure}

For completeness, we perform similar studies shown in Section~\ref{sec:prac_qrc} for a   QRC framework. The motivation for this work is to analyse how the addition of a recurrent layer affects the learnability of QRCs. As previously shown in Figure~\ref{fig:LCV_comb},  the relationship between the leak rate and the maximum CLE $\lambda_{CLE}$ is not injective (not monotonic). The main reason for this is the addition of recurrent states in the reservoir update, which changes the spectral radius of the reservoir and now the \textit{effective} spectral radius governs the rate of dissipation \citep{JAEGER2007335}. We use a similar fully connected quantum reservoir as RF-QRC with the same hyperparameters shown in Table~\ref{tab: param_RFQRC} and perform the stability analysis of the Lorenz-63 system. 
In Figure~\ref{fig:c2_LEs}, we compute the autonomous LE predictions of the Lorenz-63 system for various $\epsilon$ values, uniformly distributed within $[0,1]$. Similarly to the results of Figure~\ref{fig:c4_LEs}, for small $\epsilon$, QRCs can accurately capture the positive $\lambda_{1}$ and neutral $\lambda_{2}$ Lyapunov exponents, but cannot capture the negative $\lambda^{*}$ Lyapunov exponent. 
Furthermore, in contrast to RF-QRC, which remains conditionally stable for large $\epsilon$ values, the performance of QRCs when $\epsilon > 0.6$ ($|\lambda_{CLE}| > 100$) becomes inaccurate. The addition of recurrent connections enhances the spectral radius and reservoir memory, and when the artificial dissipation with a leak rate $\epsilon$ is not strong enough, the  QRCs becomes unstable. This analysis further informs about the design of future quantum reservoir computers, by showing that a very high expressivity without dissipation can result in an unstable reservoir. To address this issue, our proposed leaky integrated RF-QRC model, or combining other QRCs with classical leaky-integrators to tune dissipation. Thereby, helping in the design of efficient quantum reservoir computers.

\end{appendix}

\end{document}